\documentclass[aps,floats,nofootinbib,amssymb,preprint]{revtex4}

\usepackage{graphicx}

\usepackage{type1cm}
\usepackage{eso-pic}
\usepackage{color}

\newcommand{\be}{\begin{equation}}
\newcommand{\ee}{\end{equation}}
\newcommand{\bea}{\begin{eqnarray}}
\newcommand{\eea}{\end{eqnarray}}

\newcommand{\half}{\frac{1}{2}}

\newcommand{\mueff}{\mu_{\rm eff}}

\newcommand{\ba}{\begin{array}}
\newcommand{\ea}{\end{array}}
\newcommand{\bi}{\begin{itemize}}
\newcommand{\ei}{\end{itemize}}
\newcommand{\ben}{\begin{enumerate}}
\newcommand{\een}{\end{enumerate}}

\newcommand{\N}{\chi^0}
\newcommand{\C}{\chi^\pm}

\newcommand{\BM}[1]{{\mbox{\boldmath{$#1$}}}}
\begin{document}
\preprint{\font\fortssbx=cmssbx10 scaled \magstep2
\hbox to \hsize{
\hfill$\vcenter{      \hbox{\bf NSF-KITP-08-94}
\hbox{\bf MADPH-08-1512}
               \hbox{June 2008}}$}
}

\title{\vspace*{.5in}
Spin Dependence of Dark Matter Scattering}

\author{Vernon~Barger$^{a,b}$}
\email{barger@pheno.physics.wisc.edu}

\author{Wai-Yee~Keung$^{c}$}
\email{keung@uic.edu}

\author{Gabe~Shaughnessy$^{a}$}
\email{gshau@hep.wisc.edu}


\affiliation{
$^a$Department of Physics, University of Wisconsin, 1150 University Avenue, Madison, Wisconsin 53706 USA\\
$^b$Kavli Institute for Theoretical Physics, University of California, Santa Barbara, CA 93106 USA\\
$^{c}$Physics Department, University of Illinois at Chicago, Illinois 60607--7059 USA
\vspace*{.5in}}

\begin{abstract}
New experiments designed to discover a weakly interacting dark matter (DM) particle via spin dependent scattering can distinguish models of electroweak symmetry breaking.  The plane of spin dependent versus spin independent DM scattering cross sections is a powerful model diagnostic.  We detail representative predictions of mSUGRA, singlet extended SM and MSSM, a new Dirac neutrino, Littlest Higgs with $T$-parity (LHT) and Minimal Universal Extra Dimensions (mUED) models.  Of these models, the nMSSM has the largest spin dependent (SD) cross section.  It has a very light neutralino which would give lower energy nuclear recoils. The Focus Point region of mSUGRA, mUED and the right handed neutrino also predict a very large SD cross section and predict a large signal of high energy neutrinos in the IceCube experiment from annihilations of dark matter in the Sun.  We also describe a model independent treatment of the scattering of DM particles of different intrinsic spins.

\end{abstract}

\thispagestyle{empty}

\maketitle


\section{Introduction}

The nature of the dark matter (DM) that comprises about 23\% of the energy density of the Universe is unknown and a large experimental effort is devoted to its explication.  The premise is that dark matter is a stable elementary particle.  The large scale structure of the Universe excludes hot dark matter as the  primary DM component.  There are a number cold and warm dark matter particle candidates, many of which are well motivated by theoretical models of TeV scale physics. For recent reviews see Refs.~\cite{Baer:2008uu,Hooper:2008sn,Arrenberg:2008wy,dm08:2008aa,ppc08:2008aa}. Some DM candidates (e.g. axino, heavy gravitino, keV sterile neutrino) may have only cosmological consequences.  The axion, which can solve the strong CP problem, may be detectable from their conversion by photons in a strong magnetic field.  The DM particle of models with a conserved discrete symmetry (e.g. the stable particle of supersymmetry with a conserved R-parity) can have rich consequences for collider experiments, direct detection experiments  (the elastic scattering of DM particles on nuclei), and astrophysical experiments (via the detection of gamma rays, positrons, antiprotons, antideuterons, or neutrinos from dark matter annihilations in the galatic halo or the center of the Sun).   A weak scale cross section can naturally lead to a dark matter density of the requisite value to explain the WMAP determination~\cite{Komatsu:2008hk}.  Our interest here is in the tests of weakly interacting massive particles (WIMP) models via their elastic scattering on nuclei.

In WIMP models a discrete symmetry exists that makes the DM candidate
stable.   In the minimal supersymmetric standard model (MSSM), the DM particle is the lightest neutralino,
a spin 1/2 Majorana particle.  In a non-SUSY context,
a heavy Dirac fermion can be a viable DM
candidate~\cite{Boehm:2003hm,Fayet:2004bw,Adibzadeh:2008pe}.  Models such as Minimal Universal
Extra Dimensions~\cite{Antoniadis:1990ew,Randall:1999ee,Appelquist:2000nn,ArkaniHamed:1998nn,Nath:1999mw,Dicus:2000hm,Servant:2002aq,Macesanu:2005jx,Kong:2005hn,Hooper:2007qk} incorporate a Kaluza-Klein (KK) parity which
keeps the lightest KK particle  stable.  Here, the DM particle is
spin-1.  In the Littlest Higgs model solution to the gauge hierarchy problem, a discrete $T$-parity ensures that a spin-1 particle is stable~\cite{Low:2004xc,Hubisz:2004ft}.  Scalar DM, with a spin-0 field can be a consequence of singlet Higgs field extensions of the Standard Model which have been explored in a number of works~\cite{Silveira:1985rk,McDonald:1993ex,Boehm:2003hm,Burgess:2000yq,Barger:2007im,Davoudiasl:2004be,Fayet:2004bw} and a model with two extra dimensions~\cite{Dobrescu:2007ec}.  There are a variety of other models with a DM particle that we do not specifically consider here, see e.g.~\cite{Dolle:2007ce,Chang:2006ra,TuckerSmith:2004jv,Ishiwata:2007ck,Han:1997wn}

SI scattering occurs though Higgs exchange and thus may well be much smaller than SD scattering which occurs through Z-exchange~\footnote{In SUSY, sfermion exchanges can contribute to both SI and SD scattering.  However, their contributions are typically small, even for light squarks.}.  The calculation of the SI cross section has substantial uncertainties from the estimate of the Higgs coupling to strange quarks~\cite{Ellis:2005mb,Ellis:2008hf}.   The first WIMP searches focused on spin-independent (SI) scattering on a nucleus and have placed limits approaching $10^{-8}$ pb.  The SI experiments are designed to detect the coherent recoil of the nucleus caused by the DM scattering.  For a heavy nuclear target, the coherent scattering increases the cross-section by the square of the Atomic Number.   The best current limits on SI scattering have been placed by the CDMS 5 tower~\cite{Ahmed:2008eu} experiment and the XENON10 experiment~\cite{Angle:2007uj}.  The CDMS upper bound on the WIMP nucleon cross section is $4.6\times10^{-8}$ pb for a WIMP mass $\sim 60$ GeV.   The XENON10  bound is $4.5\times10^{-8}$ pb for a WIMP mass of $\sim 30$ GeV and $8.8\times10^{-8}$ pb for a WIMP mass of $\sim 60$ GeV. 

The present SI experiments have detector masses of order 10 kg.  The next generation experiments will have about 100 kg mass.  Detectors with 1000 Kg (1T) size are under design.  The XENON100 experiment is expected to reach a SI cross section sensitivity of $10^{-9}$ pb for a 100 GeV WIMP~\cite{Aprile:2008dm}.  The LUX experiment, with a 100 kg xenon detector, expects to reach $4\times10^{-10}$ pb at $M_{DM}\sim40$ GeV~\cite{Wang:2008dm}. The ultimate goal of XENON1T is sensitivity to a SI cross section of $10^{-10}$ pb.  One ton Liquid Argon detectors promise competitive sensitivities~\cite{Nikkei:2008dm,warp,Laffranchi:2007da}.

The SD cross-section occurs through the axial vector coupling to the spin content of the nucleus; there is a $J(J+1)$ enhancement from the nuclear spin $J$. Until recently, the SD cross section limits had been about 6 orders of magnitude weaker than for SI, with the best bound from ZEPLIN-II at 0.07 pb for a DM particle of mass $\sim50$ GeV scattering on neutrons~\cite{al.:2007xs}.  The NAIAD, COUPP, and KIMS experiments had placed upper bounds on the cross section for SD scattering of dark matter on protons of 0.5 pb, 0.3 pb and 0.2 pb, respectively, for a DM mass of order 100 GeV~\cite{Alner:2005kt,Bolte:2006pf,Lee.:2007qn}. The SuperKamiokande (SK) search for neutrinos from DM annihilations in the Sun placed a stronger limit, albeit model dependent, on the SD cross section;  converting the SK neutrino flux limit requires assumptions about the DM mass and the cross section for DM annihilation to neutrinos.  For a DM mass of order 100 GeV the SK limits on the SI and SD cross sections are of order $10^{-5}$ and $0.6\times10^{-2}$ pb, respectively~\cite{Habig:2001ei,Desai:2004pq}.   Liquid Xenon has about 50 percent odd isotopes which allows a measurement of SD scattering.  New SD limits  have been reported by the XENON10 experiment of $0.5\times10^{-2}$ pb for scattering on protons and $0.5$ pb for scattering on neutrons at a WIMP mass of $\sim 30$ GeV~\cite{Manalaysay:2008dm}.  Models generally predict the same cross section for scattering on protons and on neutrons, so the most restrictive of the two can be imposed in constraining models.

New techniques with bubble technologies offer great promise for exploring much smaller SD cross sections in future experiments.  The Picasso experiment in SNOLAB uses a superheated liquid in a gel.  Its current proton-WIMP SD cross section bound is of order 1 pb~\cite{BarnabeHeider:2005pg}.  In a future third stage of this experiment, with a 100 kg detector, sensitivity for SD scattering down to $10^{-4}$ pb is anticipated.  The COUPP experiment~\cite{Bolte:2006pf,Behnke:2008zz} uses a continuously sensitive bubble chamber with compounds $CF_3I$ and $C_4 F_{10}$ of fluorine, $F^{19}$, and iodine, $I^{127}$.  WIMP scattering from fluorine is determined by $\sigma_{SD}$ and scattering from iodine is determined by $\sigma_{SI}$.  

A strong SD cross section $\gtrsim 10^{-4}$ pb would also likely yield an observable signal of high energy neutrinos from the Sun due to DM annihilations~\cite{Barger:2001ur,Cirelli:2005gh,Halzen:2005ar,GonzalezGarcia:2005xw,Barger:2007hj,Barger:2007xf}.  A large signal in neutrino telescopes such as IceCube of high energy neutrinos from DM annihilations in the Sun can occur only if there is a large SD scattering (since the SI cross section is already known to be $\ll 10^{-6}$ pb.  The capture rate is
 \cite{Gould:1992xx}
\be
C_{\odot}= 3.4\times 10^{20} {\rm s}^{-1} {\rho_{local}\over 0.3 \text{ GeV/cm}^3} \left({270\text{ km/s}\over v_{local}}\right)^3\left({\sigma_{SD}^H+\sigma_{SI}^H + 0.07 \sigma_{SI}^{He}\over 10^{-6}\text{ pb}}\right)\left({100\text{ GeV}\over m_{\N_1}}\right)^2,
\label{eq:caprate}
\ee
where $\rho_{local}$ and $v_{local}$ are the local density and velocity of relic dark matter, respectively.  Present limits on the SI and SD scattering cross section on nuclei strongly suggest that an observation of these neutrinos from the Sun would only result from SD capture of WIMPS by hydrogen in the Sun.   To calculate the expected signal rates, we closely follow the approach of Ref.~\cite{Barger:2007xf}.

Recently, the DAMA/LIBRA experiment and the predecessor DAMA/NaI experiment have reported a combined 8.2$\sigma$ evidence for an annual modulation signature for DM particles in the galactic halo~\cite{Bernabei:2008yi}.  The implied SI cross section would be about $10^{-6}$ pb. A possible way to reconcile the tension with other more restrictive bounds from other experiments is that DAMA is not detecting nuclear recoils, but electromagnetic energy deposition from axion-like interactions in the detector.  For now, we set aside this interesting experimental development and focus on nuclear recoil detection of DM.

Our study is devoted to the analysis of what can be learned from the combined measurements of SI and SD processes, which can provide a powerful diagnostic in differentiating models.  We specifically detail representative predictions of mSUGRA~\footnote{Recently, efforts have been made to distinguish specific areas of mSUGRA parameter space that yield similar collider signatures using measurements from direct detection experiments~\cite{Altunkaynak:2008ry}.}, singlet extensions of the SM and the MSSM, a new Dirac neutrino, Littlest Higgs with $T$-parity, and Minimal Universal Extra Dimensions models (mUED).  We assume that the DM particle is in thermal equilibrium in the early universe and we impose the requirement that the relic density of DM reproduces the measured WMAP density within $\pm 10\%$.  

We also present general formulas for SI and SD cross sections for the scattering of WIMPS of intrinsic spin 0, 1/2, 1, and 3/2.

The organization of the paper is as follows.  In section \ref{sect:scan}, we outline the scan proceedure while in section \ref{sect:predict} we elucidate the DM scattering predictions of representative models that provide a WIMP candidate and detail how the models could be distinguished by measurement of SI and SD scattering.  Then we proceed, in section \ref{sect:modindep}, to a model independent analysis of DM scattering for DM particles of arbitrary spin. Finally, we summarize the results of our study in  section \ref{sect:concl}.

\section{Scan Technique}
\label{sect:scan}

To arrive at the model predictions, we use a Markov Chain Monte Carlo (MCMC) approach that is based on the Bayesian methods to scan over the relevant model parameters.  With a MCMC, a random point, $x_{i}$ in parameter space is chosen with a likelihood, ${\cal L}_{i}$, assigned to it based on the associated model predictions such as the DM scattering cross sections and relic density.  We then choose another random point, $x_{i+1}$, and compare its likelihood with the previous point in the chain.  If the likelihoods satisfy
\be
{{\mathcal L_{i+1}}\over{\mathcal L_{i}}} > \zeta,
\ee
where $\zeta$ is chosen from a uniform distribution on the unit interval, the point $x_{i+1}$ is accepted and appended to the chain, otherwise a copy of $x_{i}$ is appended to the chain.  

The power of the MCMC approach is that in the limit of large chain length, the distribution of points, $x_{i}$ approach the posterior distribution of model parameters for the constraining data.  The speed at which the parameter space is scanned by MCMC is generally faster than traditional random or grid scans, with the difference in speed between the two methods increasing with larger parameter dimension.  

We construct the likelihood from observables which have a central value, $\mu_{j}$, with uncertainty, $\sigma_{j}$, through the corresponding $\chi^{2}$ value
\be
{\mathcal L_{i}} = e^{-\sum_{j}\chi_{j}^{2}/2}=e^{-\sum_{j}(d_{ij}-\mu_{j})^{2}/2\sigma_{j}^{2}},
\ee
where $d_{ij}$ is the model prediction for the $j^{\rm th}$ observable of the constrained data based on the parameters in the chain $x_{i}$.  To impose exclusion limits, we follow the approach of Ref.~\cite{Roszkowski:2007fd}.  

To construct the posterior distribution, we obtain the covariance matrix among the relevant parameters of the first 200 unique links in the MCMC chain and discard them as a ``burn-in''.  We define the ranges over which the parameters are allowed to vary later in Section \ref{sect:predict} for each model.  We use this distribution information to construct more efficient chain proposals  for subsequent links in the new chain as outlined in Ref.~\cite{Baltz:2006fm}.

Convergence of the chain is tested using the method of Raferty and Lewis~\cite{Raferty:1992aa,Raferty:1992ab,Raferty:1995aa}.  In checking that the chain has converged, we require that the posterior distribution is within 2\% of the 95\% quantile with 95\% C.L.  This typically results in a few thousand unique links in the chain.  We then double check convergence using the method of Ref. ~\cite{Dunkley:2004sv}.  For more extensive reviews of this type of approach to parameter estimation, see Refs.~\cite{deAustri:2006pe,Dunkley:2004sv,Roszkowski:2007fd,Baltz:2006fm}. 

\section{Predictions of Models for DM scattering}
\label{sect:predict}

The measured DM density from the WMAP5 analysis of cosmological data (cosmic microwave background radiation data combined with distance measurements from Type Ia supernovae and baryon acoustic oscillations in the distribution of galaxies)~\cite{Komatsu:2008hk} is
\be
\Omega_{\rm DM} h^2=0.1143\pm 0.0034,  \quad \quad h=0.701\pm 0.013
\ee

We scan over the parameters of models with the constraint that the models give this relic density within $\pm 10\%$ assigned uncertainty associated with combined galactic and particle physics uncertainties.  We separately consider model parameters that give a DM density below the measured value to allow for the possibility that there may be more than one contributing DM particle.
Our calculations of SI and SD cross sections and the relic density are made using the Micromegas 2.1 code~\cite{Belanger:2008sj} with an isothermal dark matter density profile and a relative DM velocity in the galactic halo of $v_{DM}=220$ km/s.  In calculating the SI scattering cross sections we adopt the recent determinations of the pion-nucleon sigma term $\sigma_{\pi N}=55$ MeV and the sigma term that determines the mass shift of the nucleon due to chiral symmetry breaking $\sigma_{0}=35$ MeV~\cite{Belanger:2001fz,Belanger:2004yn,Belanger:2006is,Belanger:2008sj}.

\subsection{mSUGRA model}

Supersymmetry (SUSY) has a conserved $R$-parity that renders the lightest supersymmetric particle stable. Supersymmetry stabilizes the radiative corrections to the Higgs boson mass and realizes Grand Unification of the electroweak and strong couplings.  Moreover, the quantum numbers of the Standard Model particles are explained by a singlet ${\bf 16}$ representation of $SO(10)$ for each generation of fermions.  The necessary breaking of the supersymmetry is  achieved in the minimal supergravity model (mSUGRA) by Planck scale mediation between the observed and hidden sectors of the theory.  The mSUGRA model has become a reference standard~\cite{Baer:2006rs,Binetruy:2006ad,Drees:2004jm}.  It has only a small number of parameters $(m_0, m_{1/2}, A_0, \tan\beta, sign(\mu))$\footnote{The common scalar, gaugino masses and the soft trilinear terms are unified at the GUT scale to chosen values of $m_{0}$, $m_{1/2}$ and $A_{0}$, respectively.  After specifying the ratio of the two Higgs VEVs, $\tan \beta = {\langle H_{u}\rangle \over \langle H_{d}\rangle }$ and the sign of the Higgsino mass parameter, $\mu$, the model at the electroweak scale is fully defined by the GUT scale parameters through the RGE running. }.  There are four regions of mSUGRA parameter space that give values of the relic density at or below the WMAP value.  These are:
\bi

\item Focus Point Region (FP)~\cite{Baer:1995nq,Chan:1997bi,Feng:1999zg}, also called the Hyperbolic Branch:  This region is preferred by $b-\tau$ unification and new experimental results from $b\to s \gamma$~\cite{Roszkowski:2007fd}. The SUSY flavor changing neutral current and CP violating problems are resolved naturally by large sfermion masses.  At large $m_0$, the superpotential Higgsino mass term $\mu$ becomes quite small, and the lightest neutralino, $\N_1$, is a  mixed higgsino-bino state.  Neutralino annihilation in the early universe to vector bosons is enhanced.  This results in an enhanced SD cross section and enhanced neutralino annihilation in the Sun to neutrinos.  The FP region would allow a precision gluino mass measurement ($\pm8$\%) for $M_{\tilde g}$ in the range of 700 to 1300 GeV~\cite{Baer:2007ya}.  The FP region falls into a more general class of neutralino models which are termed ``Well-tempered Neutralinos'' in which the lightest neutralino is a Bino-Wino or Bino-Higgsino mixture~\cite{ArkaniHamed:2006mb,Baer:2006te}.

\item  A-Funnel Region (AF):  The A-funnel region occurs at large values of the parameter $\tan \beta\approx 50$, near $2 M_{\N_1} \sim m_A$.  The neutralinos annihilate through the broad pseudoscalar Higgs resonance $A$~\cite{Drees:1992am,Baer:1995nc,Baer:1997ai}.
There is also a light Higgs resonance region where $2 M_{\N_1} \sim m_h$ at low $m_{1/2}$ values~\cite{Nath:1992ty,Baer:1995nc,Barger:1997kb}.

\item Coannihilation Regions (CA): The neutralino-stau co-annihilation region occurs at very low $m_0$ but any $m_{1/2}$ values, so that
$M_{\tilde \ell}\sim M_{\N_1}$, and neutralinos can annihilate against tau sleptons~\cite{Ellis:1998kh,Ellis:1999mm} in the early universe.
For certain $A_0$ values which dial $m_{\tilde t_1}$ to very low values, there also exists a stop-neutralino co-annihilation region~\cite{Ellis:2001nx}.

\item Bulk Region (BR): The bulk region is at low $m_0$ and low $m_{1/2}$, where neutralino annihilation is enhanced by
light t-channel slepton exchange~\cite{Baer:1995nc,Barger:1997kb}. The WMAP determination of the relic density has pushed this
allowed region to very small $m_0$ and $m_{1/2}$ values, while LEP2 limits on $M_{\C_1}$
and $m_{h}$ exclude these same low values so that the bulk region is disfavored~\cite{Ellis:2003cw,Baer:2003yh,Djouadi:2006be}.  
\ei

\begin{figure}[t]
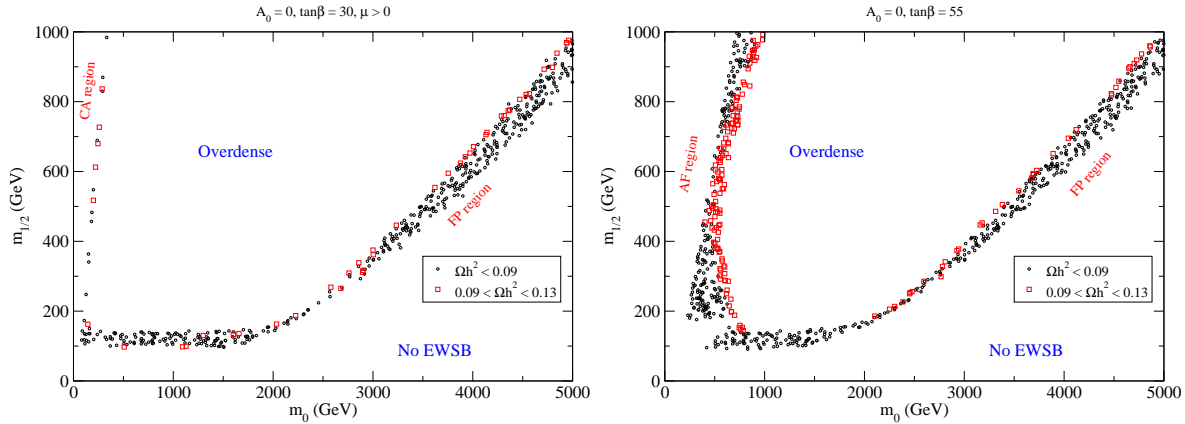

\begin{center}
      \includegraphics[angle=0,width=0.47\textwidth]{figs/msugscan-tb-30.eps}
      \includegraphics[angle=0,width=0.47\textwidth]{figs/msugscan-tb-55.eps}
\caption{Scan over the common scalar, $m_{0}$, and gaugino mass, $m_{1/2}$, to satisfy relic density and LEP2 constraints on mSUGRA with specific values of $A_{0}$ and $\tan \beta$ ($A_{0}=0$, $\tan\beta=30, 55$).  Open squares (in red) show parameter values that have a relic density within $\pm 10\%$ of the measured value; solid points (in black) have lower relic density values.}
\label{fig:msugscan}
\end{center}
\end{figure}

We note that neutrino Yukawa couplings can significantly affect the RGE evolution and the calculated neutralino relic density in regions of parameter space where soft SUSY-breaking slepton masses and/or trilinear couplings are large~\cite{Barger:2008nd}.  The changes can be large in the focus point, A-funnel, and stop-coannihilation regions of mSUGRA and can expand the allowed regions of the mSUGRA parameter space and the dark matter direct and indirect detection rates.  We do not fold in the uncertainties associated with the neutrino couplings in the present study.

\begin{figure}[t]
\begin{center}
      \includegraphics[angle=0,width=0.47\textwidth]{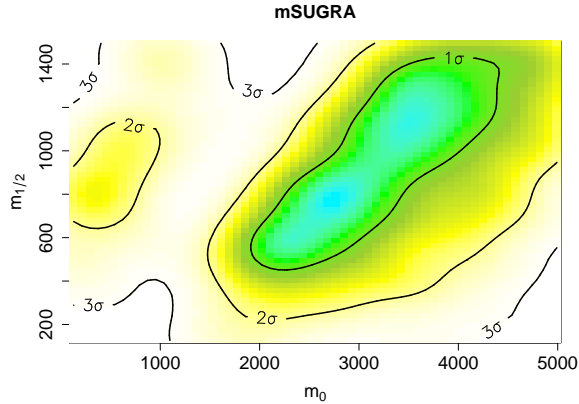}
\caption{Neutralino relic density and LEP2 allowed regions in the mSUGRA model obtained from a scan over the common scalar, $m_{0}$, and gaugino mass, $m_{1/2}$, while allowing variations of the parameters $A_{0}, \tan \beta$ and $m_{t}$.  One, two and three sigma contours are shown}
\label{fig:msugmcmc}
\end{center}
\end{figure}

Figure \ref{fig:msugscan} shows representative regions in mSUGRA parameter space where the relic density is accounted for within $\pm 10\%$ of the WMAP value (red open points) or is less than the WMAP value (black solid points).  The choices of $A_{0}=0$, $\mu> 0$, $m_{t}=170.9$ GeV and $\tan\beta = 30$ (Fig 1a) and $\tan\beta= 55$ (Fig 1b) are chosen to illustrate the FP, CA and AF regions.  Well delineated regions  that satisfy the relic density are carved out of the parameter space.  As noted above, these regions become wider when the unknown masses of the right-hand neutrinos are taken into account in the evolution from the GUT scale~\cite{Barger:2008nd}.

When we allow the trilinear parameter, $A_{0}$, and $\tan \beta$ to vary, as well as the input value of the top quark mass $m_{t}$, we arrive at a broader range of parameter values that reproduce the DM density.  In our analysis, we allow the MCMC to scan within the parameter ranges
\be
\begin{array}{ccccc}
50\text{ GeV}&\le& m_{0}& \le& 5\text{ TeV}\\
150\text{ GeV}&\le& m_{1/2}& \le& 1.5\text{ TeV}\\
1&\le& \tan \beta&\le& 50\\
-3\text{ TeV}&\le& A_{0}& \le& 3\text{ TeV}\\
0&<& \mu& & \\
\label{eq:sugranges}
\end{array}
\ee
We allow a gaussian variation of the top quark mass around its central value of $m_{t}=170.9$ GeV with a standard deviation of $1.8$ GeV~\cite{CDFD0:2007bxa}.  The marginalization over the ``nuisance parameters'' allows the uncertainty of the input parameters to be reflected in  the posterior distributions of the model parameters.  In mSUGRA, the top quark mass variation strongly affects the running from the GUT scale to the weak scale.  Therefore, through its variation, the top mass uncertainty expands the parameter ranges that are consistent with the observed relic density compared to results obtained by fixing the top mass as is often done in such studies.  The resulting distribution of the common scalar and gaugino masses ($m_{0}, m_{1/2}$) is shown in Fig. \ref{fig:msugmcmc}.  

\begin{figure}[t]
\begin{center}
\includegraphics[angle=0,width=.47\textwidth]{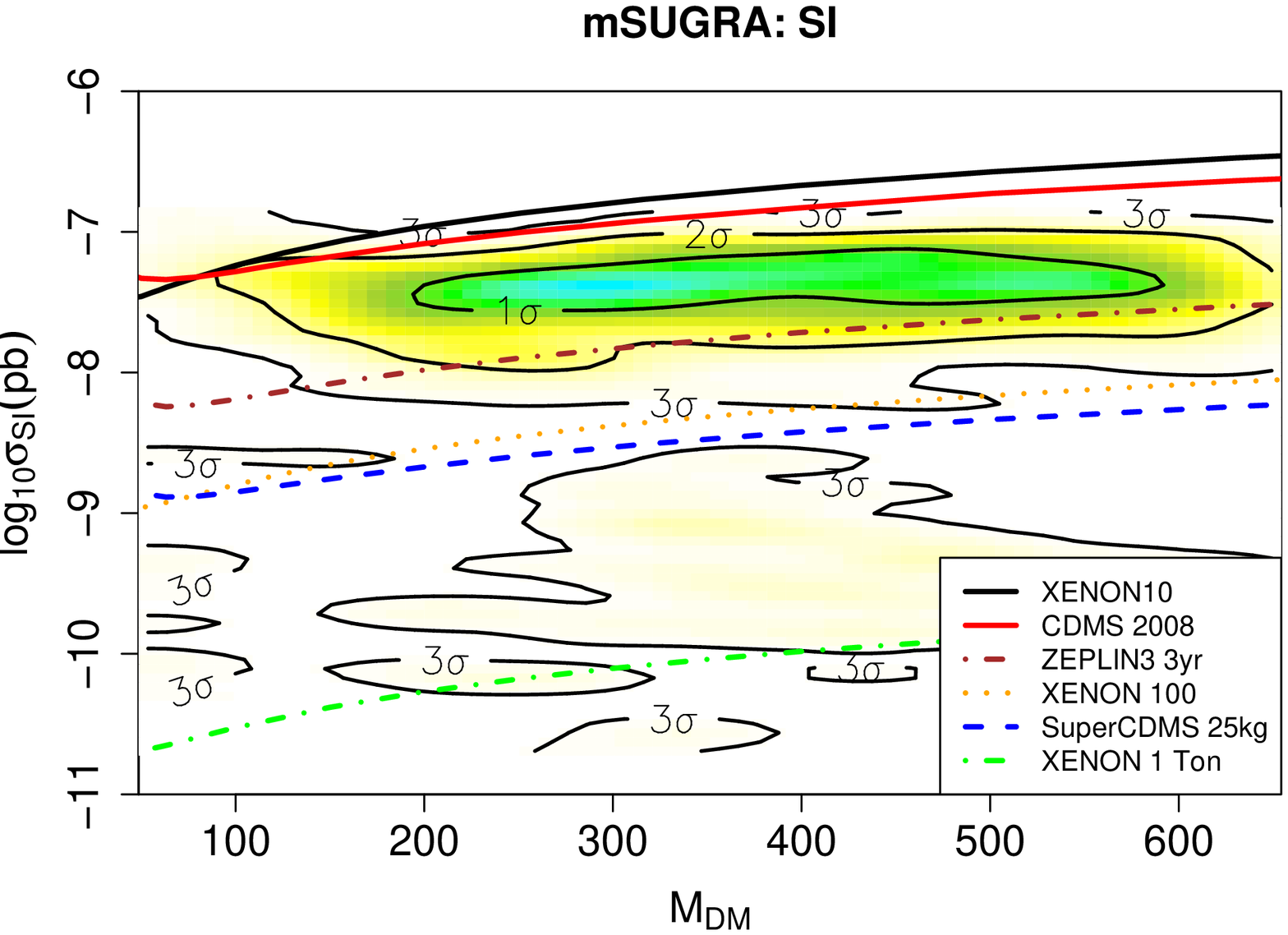}
\includegraphics[angle=0,width=.47\textwidth]{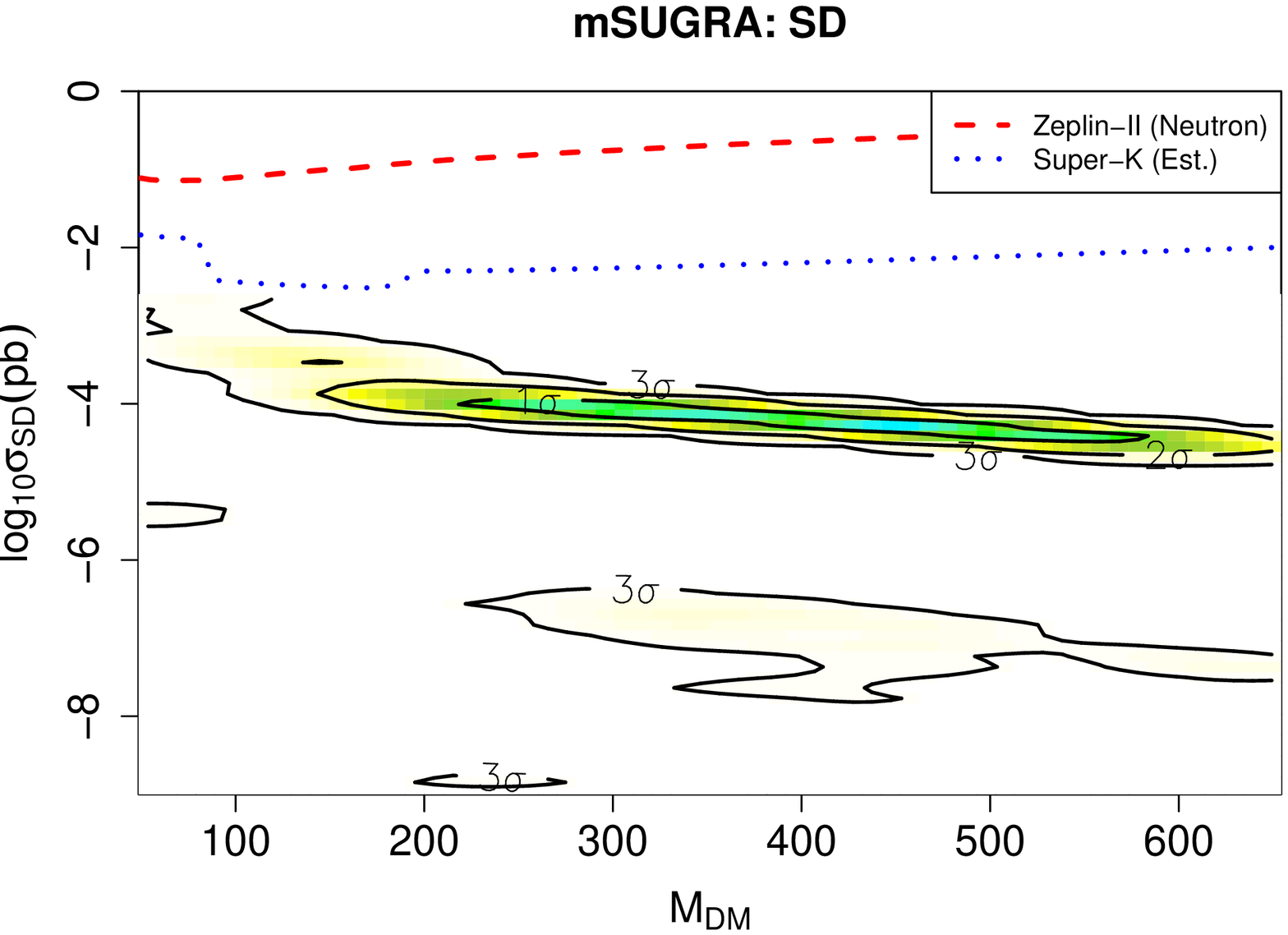}
\includegraphics[angle=0,width=.6\textwidth]{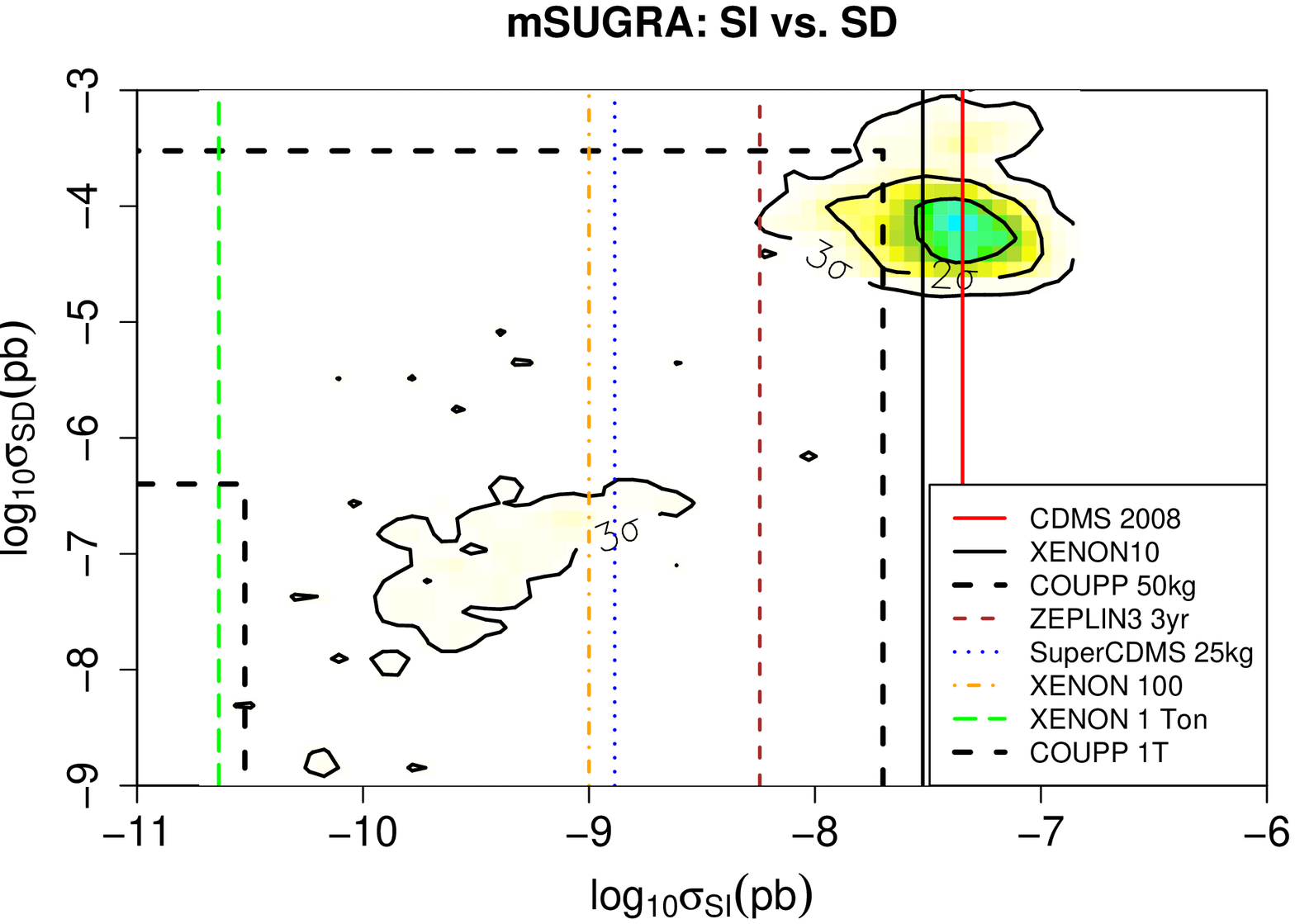}
\caption{Posterior distributions of SI and SD cross sections in the mSUGRA model.  The region of high SI and SD cross sections corresponds to the FP region and should be probed fully by the XENON10 and Super CDMS 25 Kg experiments.  The tail that extends to lower SI and SD values corresponds to the AF and CA regions.  The proposed COUPP1T experiment should probe these regions~\cite{Bertone:2007xj}.}
\label{fig:sugsipsdp}
\end{center}
\end{figure}

In general the neutralino composition in terms of the gaugino ($\tilde B$, $\tilde W^{3}$) and Higgsino ($\tilde H_1, \tilde H_2$) states is given by
\be
 \chi_1^0 = N_{11}\tilde B+N_{12}\tilde W^{3}+N_{13}\tilde H_{1}+N_{14} \tilde H_{2}
\ee
The spin dependent scattering cross section is largely governed by $Z$-boson exchange and is sensitive to the Higgsino asymmetry
\be
\sigma_{SD}\propto |N_{13}^2-N_{14}^2|^2
\ee
The FP region is unique among the mSUGRA regions that reproduce the relic density in having a large Higgsino asymmetry and a correspondingly large $\sigma_{SD}$

The SD and SI scattering cross sections in mSUGRA are displayed in Fig. \ref{fig:sugsipsdp}, where it is apparent that different solutions to the DM relic density populate different regions of $\sigma_{SD}$ versus $\sigma_{SI}$.  The limit of current DM recoil experiments is strongest in the mass range $M_{DM}\sim 50-100$ GeV.  In all subsequent $\sigma_{SI}$ vs. $\sigma_{SD}$ results, we present the current or projected best limit of the recoil experiments.  The use of $\sigma_{SD}/\sigma_{SI}$ to distinguish mSUGRA regions has also been advocated previously, see e.g. Ref.~\cite{Bertone:2007xj}.  However, it should be noted that the experimental limits are progressively less constraining at higher and lower $M_{DM}$ than the best limits.  The FP region with its relatively large $\sigma_{SD}$ can be definitively tested by SD, SI measurements.  DM detection corresponding to the FP region would have major significance for colliders in that high mass sfermions would be implied.  Similar to the spin-0 DM case, detection of SD scattering of DM would immediately rule out a model with a gaugino dominated neutralino~\cite{Acharya:2008zi} which predicts a vanishingly small SD cross section.

\begin{figure}[htbp]
\begin{center}
      \includegraphics[angle=0,width=0.57\textwidth]{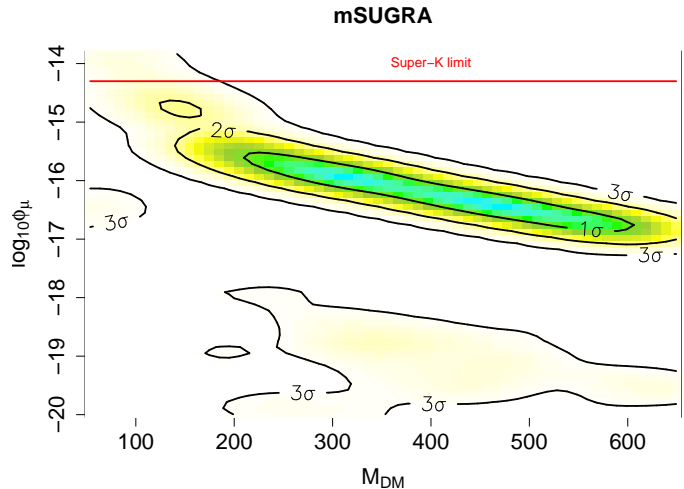}
\caption{Posterior distributions of the induced muon flux at the Super-Kamiokande detector (cm$^{-2}$s$^{-1}$ units) from high energy neutrinos created by DM annihilations in the Sun for mSUGRA.  Only the low neutralino mass FP region is constrained by the Super-K limit.  }
\label{fig:superk}
\end{center}
\end{figure}

The Super-Kamiokande experimental upper bound on the induced muon flux is $5\times 10^{-15}$ cm$^{-2}$ s$^{-1}$~\cite{Desai:2004pq}, and is shown by the horizontal line in Fig. \ref{fig:superk}.  This bound only eliminates only a small fraction of the FP parameter space.

\subsection{Singlet Higgs Extensions of the MSSM}

The motivation for appending a singlet Higgs field $S$ to the MSSM is to explain why the Higgsino mixing parameter $\mu$ in the superpotential
\be
W=\mu \hat H_u \cdot \hat H_d 
\ee
is at the TeV scale.  The dynamical $\mu$ solution is to let the $\mu$ parameter be generated by a vacuum expectation value of a S of order $M_{SUSY}$.
\be
\mueff=\lambda \langle S\rangle
\ee
Three classes of singlet model extensions xMSSM have been considered~\cite{Panagiotakopoulos:1998yw,Dedes:2000jp,Choi:2004zx,Gunion:2005rw,Barger:2006dh,Barger:2006kt,Barger:2006sk,Accomando:2006ga,Ferrer:2006hy,Barger:2007nv,Balazs:2007pf,Lee:2007fw} with additional Higgs-sector superpotential terms as follows:
\ben
\item NMSSM: $S^3$ (cubic)
\item nMSSM: $S$ (tadpole)
\item UMSSM: $U(1)'$ symmetry
\een
In the xMSSM the spin-0 S field mixes with the neutral Higgs fields of the MSSM; the spin-1/2 singlino field $\tilde S$ in the composition of the lightest neutralino mixes with the neutralinos of the MSSM.  The consequence for DM is that the $\tilde S$ component changes the DM couplings and cross sections~\cite{Barger:2006kt,Barger:2007nv}.  The predictions of the tadpole model are the most different from the MSSM~\cite{Barger:2006kt,Barger:2007nv,Balazs:2007pf}; we therefore focus on it here.  The singlet scalar field also affects LHC phenomenology as the $S$ admixture in the Higgs states changes the Higgs production and decay rates and modifies the cascade decay chains of the sparticles~\cite{Ellwanger:2004gz,Gunion:2005rw,Chang:2005ht,Barger:2006kt,Barger:2006dh,Barger:2006sk,Balazs:2007pf}.

We scan over the parameters directly relevant to the DM phenomenology of this model.  Their ranges are chosen to be
\be
\begin{array}{ccccc}
1&\le& \tan \beta&\le& 20\\
200\text{ GeV}&\le& \langle S\rangle &\le& 2\text{ TeV}\\
100\text{ GeV}&\le& \mu_{eff}& \le& 1\text{ TeV}\\
50\text{ GeV}&\le&M_{1}&\le&1\text{ TeV}\\
100\text{ GeV}&\le&M_{2}&\le&1\text{ TeV}\\
100\text{ GeV}&\le&A_{\lambda}&\le&1\text{ TeV}\\
100\text{ GeV}&\le&A_{\lambda}&\le&1\text{ TeV}\\
-(400\text{ GeV})^{2}&\le&t_{F}&\le&(400\text{ GeV})^{2}\\
-(200\text{ GeV})^{3}&\le&t_{S}&\le&(100\text{ GeV})^{3}\\
\label{eq:tadranges}
\end{array}
\ee
where the parameters $A_{\lambda}, t_{F}$ and $t_{S}$ are defined in Ref.~\cite{Barger:2006kt}.
We also place a minimum cut on the lightest neutralino mass of 10 GeV since the current DM scattering experiments have considerably reduced sensitivity to a DM mass this low.

\begin{figure}[htbp]
\begin{center}
\includegraphics[angle=0,width=.47\textwidth]{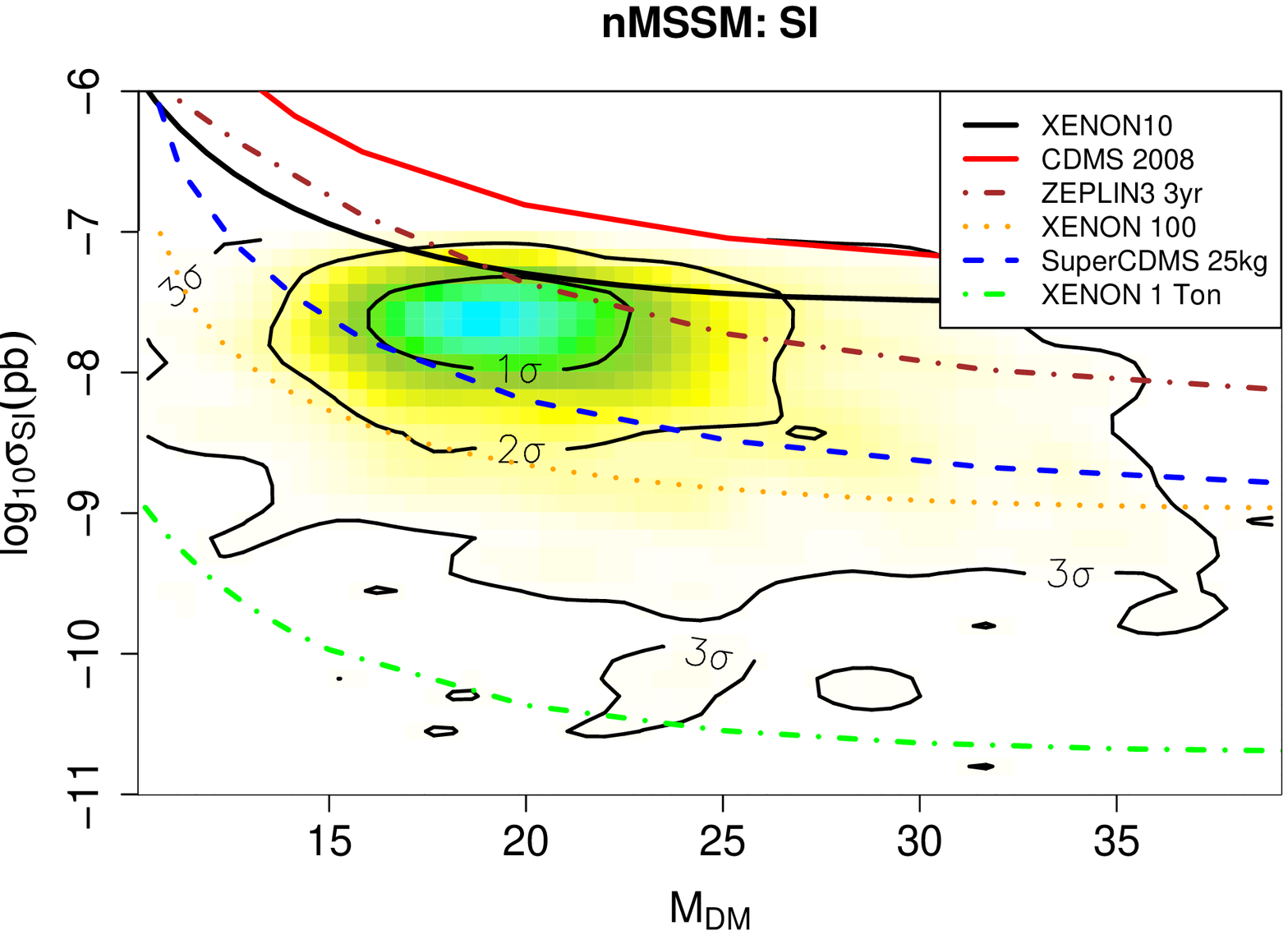}
\includegraphics[angle=0,width=.47\textwidth]{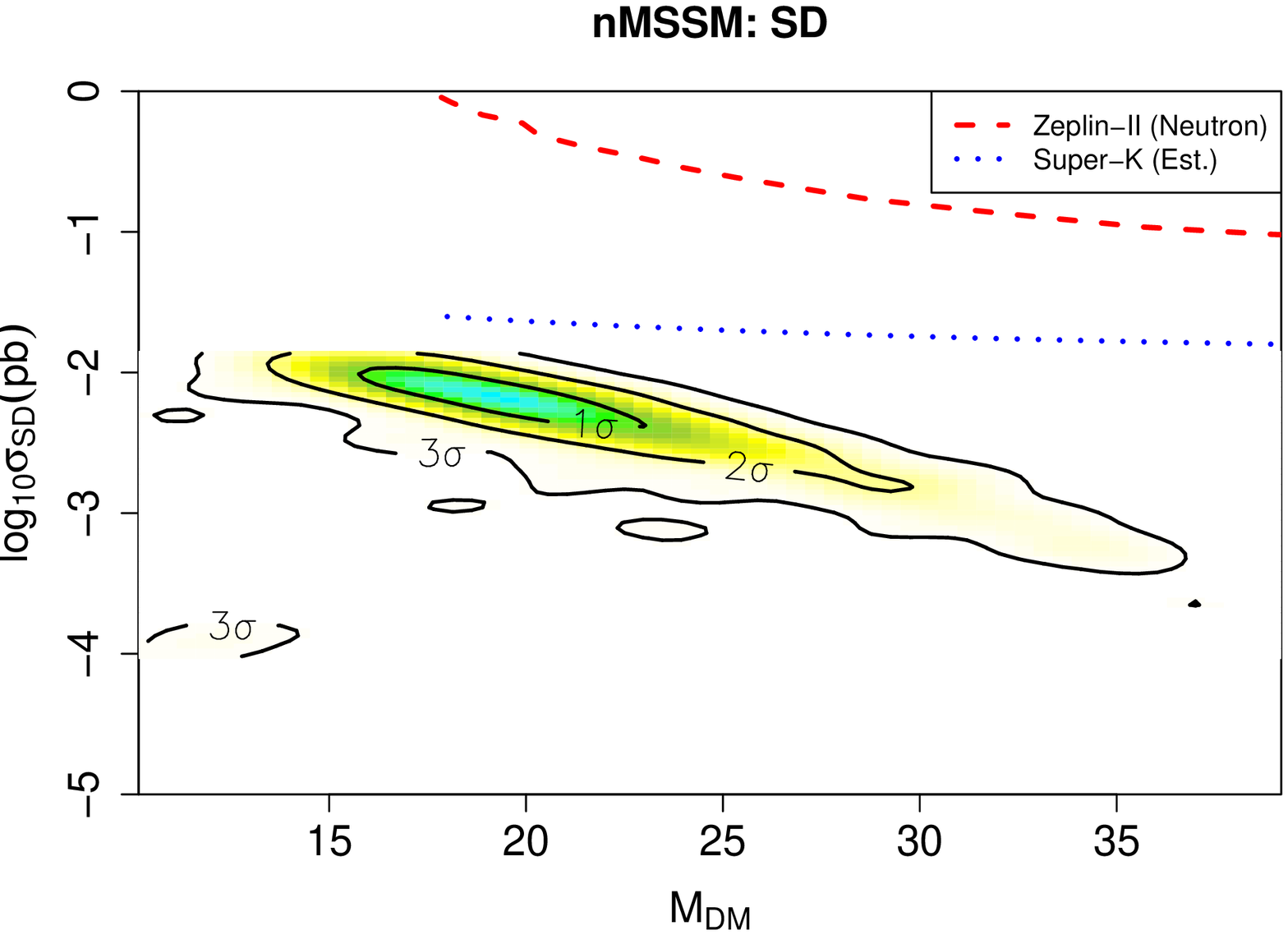}
      \includegraphics[angle=0,width=0.47\textwidth]{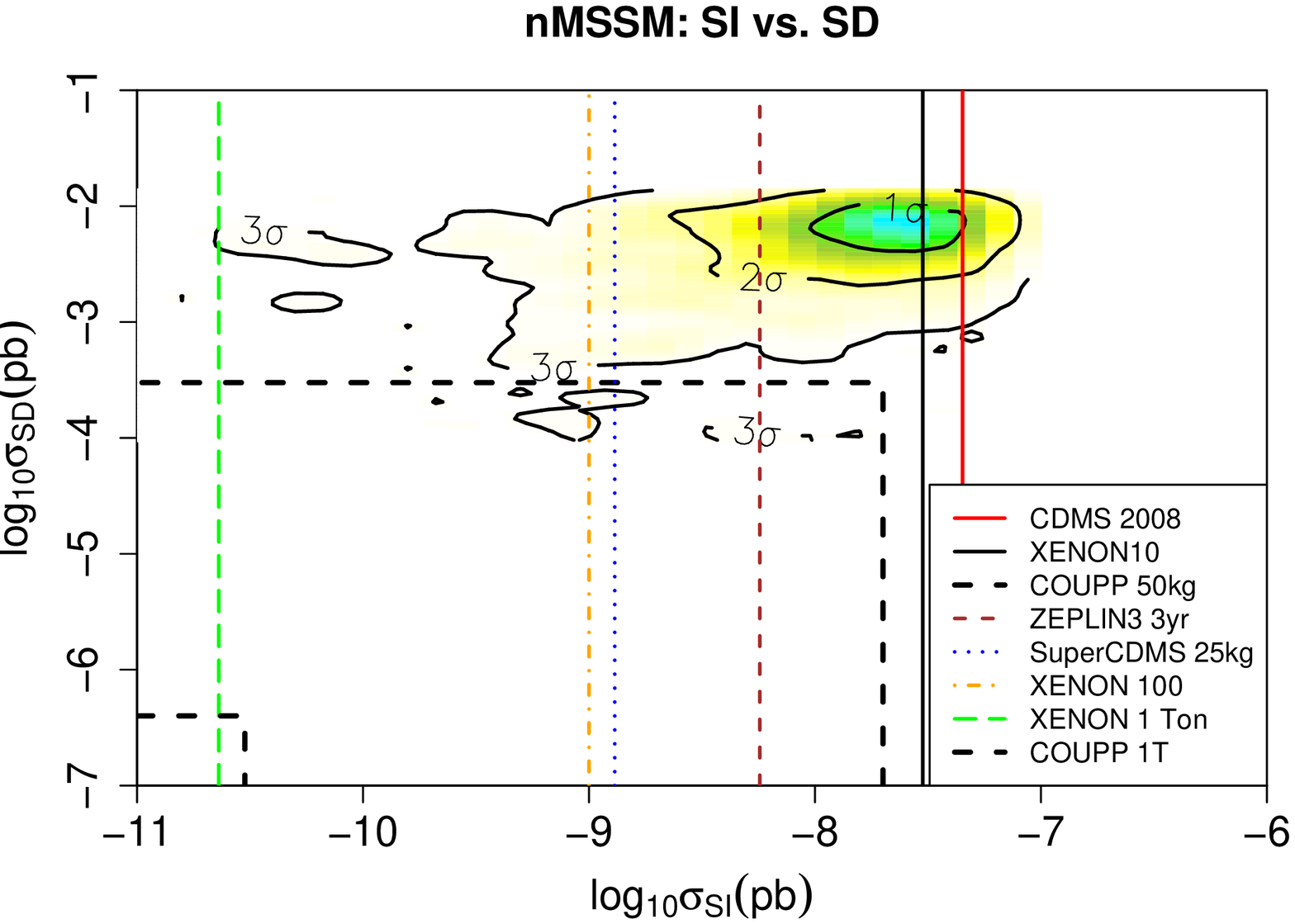}
\caption{Posterior distributions of SI and SD cross sections that satisfy the relic density and LEP2 constraints in the tadpole extended SUSY model.  This model has light neutralinos with relatively large SI and SD cross sections that make the prospects for discovery promising.}
\label{fig:xmssm}
\end{center}
\end{figure}

Figure \ref{fig:xmssm} shows the SI and SD cross sections for the nMSSM.  The contours show the posterior distributions with the present SI and SD experimental constraints enforced.  The SD cross section can be rather large since the dominant annihilation occurs through the $Z$ boson.  Therefore, to counter the small annihilation rate in the early universe due to the small neutralino mass, the neutralino pair must have a larger $Z$ boson coupling.  

The nMSSM predictions of $\sigma_{SI}$ and $\sigma_{SD}$ will be largely probed by future experiments, provided that the neutralino mass is not too small.  Relaxation of the 10 GeV lower cutoff on the DM mass would be to loosen the prospects of DM detection in this model.  

\subsection{Scalar singlet extended SM (xSM)}

One of the simplest models that includes a viable DM candidate is a real singlet Higgs extended SM with a discrete ${\mathbb Z}_{2}$ symmetry~\cite{Barger:2007im,McDonald:1993ex,Burgess:2000yq}.  In this model, a stable singlet is appended to the SM and interacts with the SM fields only through the Higgs boson.  As a consequence, the relic density and elastic scattering predictions are correlated with the Higgs sector.  The potential for a stable singlet Higgs boson, $S$, is given by
\bea
V &=& {m^2\over 2} H^\dagger H+{\lambda\over 4}(H^\dagger H)^2+{\delta_2\over 2} H^\dagger H S^2 +{\kappa_2 \over 2} S^2+{\kappa_4 \over 4} S^4,
\label{eqn:hpot}
\eea
where $H$ is the $SU(2)$ doublet field and $m^2$,  $\lambda$ are the usual SM parameters of the Higgs potential.  Combining this singlet extended Higgs sector with the rest of the SM gives the \lq\lq xSM", the extended Standard Model.  The singlet DM mass is determined by $M_{DM}^{2}=\half \kappa_{2}+{1\over 4} \delta_{2}v^{2}$, where $v=246$ GeV is the SM Higgs VEV.  In the MCMC scan, we allow the parameters to vary within the ranges 
\be
\begin{array}{ccccc}
114\text{ GeV}&\le& M_{h}&\le& 150\text{ GeV}\\
50\text{ GeV}&\le& M_{DM}&\le& 1\text{ TeV}\\
0&\le& \delta_2& \le& 4\\
0&\le&\kappa_4&\le&4
\label{eq:rsxranges}
\end{array}
\ee
where the upper limit on the Higgs mass comes from electroweak precision measurements~\cite{Barger:2007im}.

Further extensions of this model include making the singlet a complex field~\cite{Barger:2008aa} or adding right handed neutrinos, which would explain neutrino mass generation when the singlet obtains a VEV~\cite{McDonald:2007ka}.  The real singlet Higgs field has also been shown to enhance the first order phase transition of electroweak symmetry breaking that is necessary for electrweak baryogenesis and may reduce the tension between electroweak precision observables~\cite{Profumo:2007wc} and the non-observation of the SM Higgs boson below 114 GeV~\cite{Barger:2007im}.

\begin{figure}[htbp]
\begin{center}
      \includegraphics[angle=0,width=0.47\textwidth]{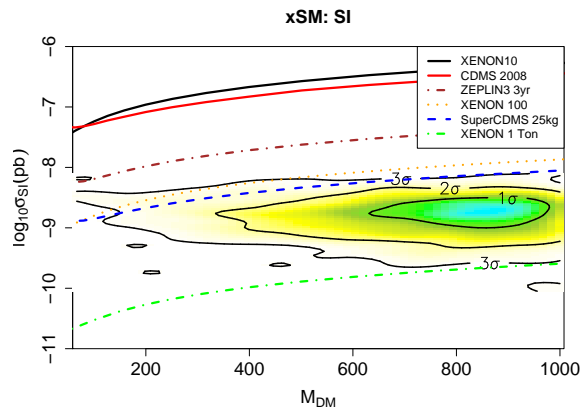}
\caption{Real singlet extended SM posterior distributions of the SI cross section vs. $M_{DM}$.}
\label{fig:rsxsip}
\end{center}
\end{figure}

Since the DM candidate in the xSM class of models is spin-0, the SD cross section vanishes.  If a SD signal is seen, this class of models would be ruled out.  The SI cross section is generally small, below $\sim 10^{-8}$ pb as seen in Fig. \ref{fig:rsxsip}.  

The xSM scenario could be indicated by an observation of a SI cross section of ${\cal O}(10^{-9}\text{ pb})$ with next generation detectors and a null result in searches for SD scattering and high energy neutrinos from the Sun in neutrino telescopes.  
\subsection{Dirac neutrino DM}

\begin{figure}[htbp]
\begin{center}
\includegraphics[angle=0,width=.47\textwidth]{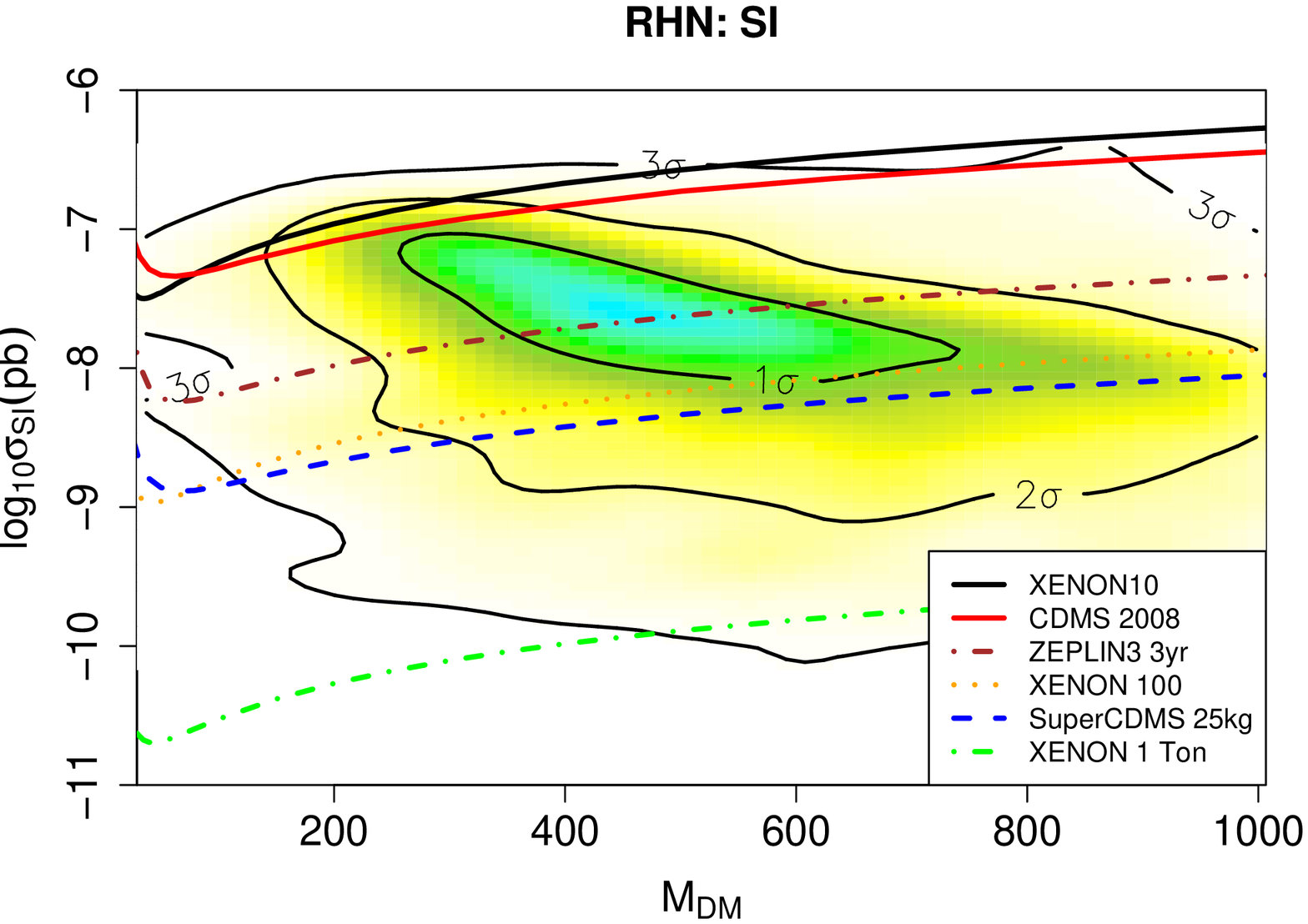}
\includegraphics[angle=0,width=.47\textwidth]{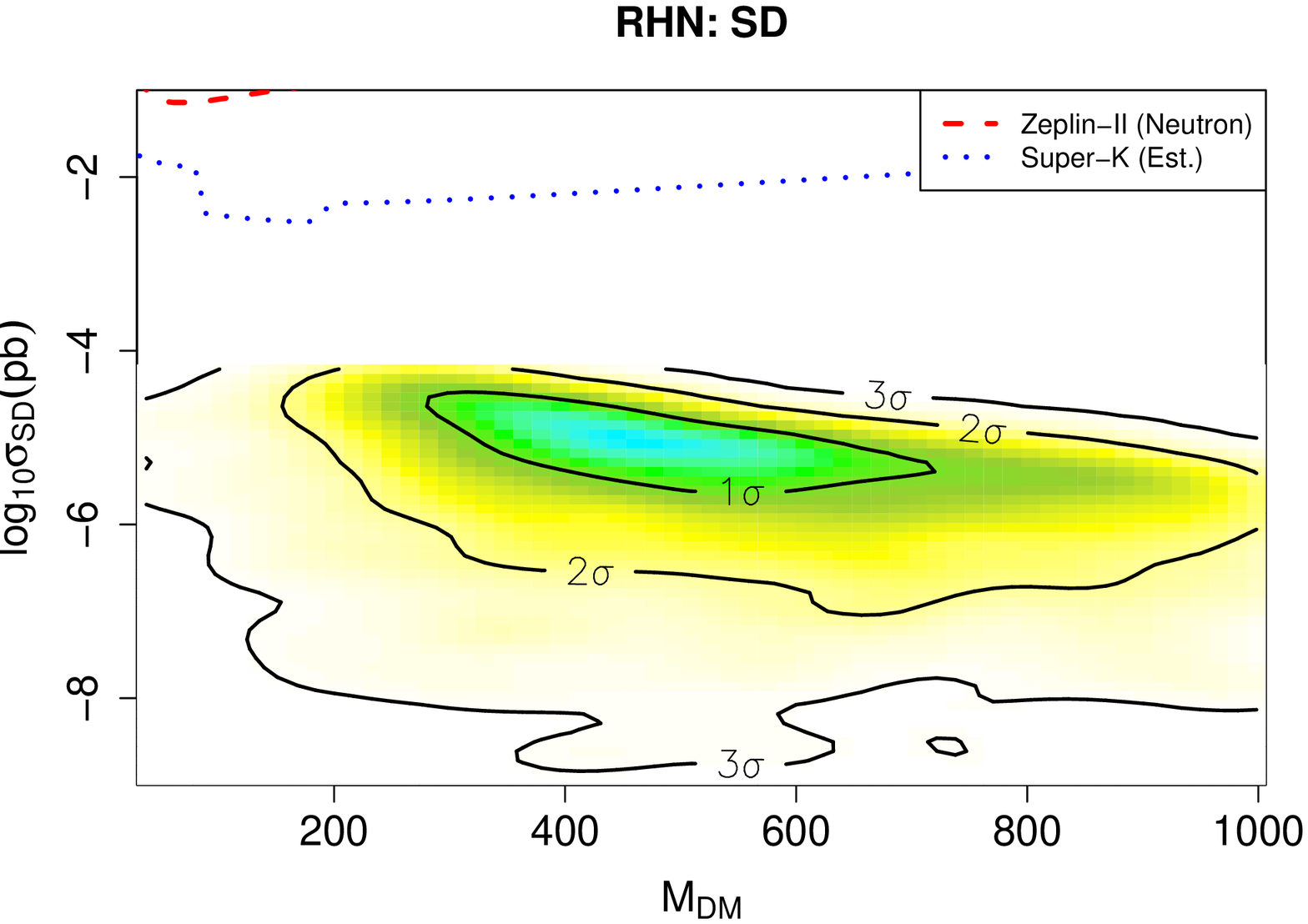}
      \includegraphics[angle=0,width=0.47\textwidth]{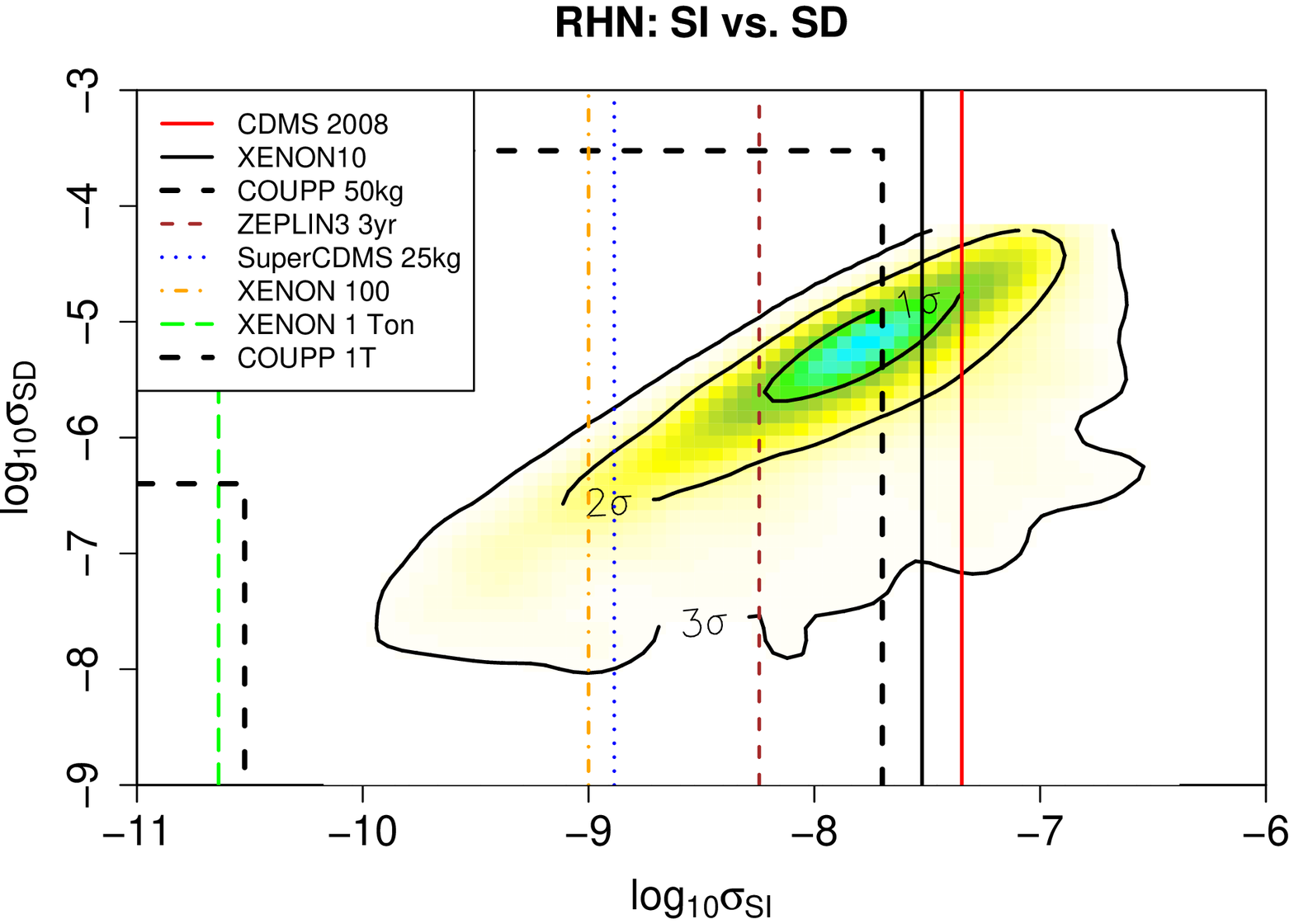}
\caption{Posterior distributions of SI and SD cross sections in the Dirac neutrino model of Ref.~\cite{Belanger:2007dx}.  This model has a heavy stable neutrino that annihilates predominantly through $Z$ and $Z'$ gauge bosons. }
\label{fig:rhn}
\end{center}
\end{figure}

A simple DM model can be realized with a stable neutral heavy Dirac fermion~\cite{Belanger:2007dx,Hooper:2005fj}.  The stability of the heavy neutrino can be guaranteed by imposing baryon number conservation in a warped GUT model.  In this model, the heavy neutrino can be ${\cal O}(100\text{ GeV})$ and annihilate in the early universe via the $Z, Z'$ and $h$ bosons.  The parameter ranges over which we allow the MCMC to scan are
\be
\begin{array}{ccccc}
114\text{ GeV}&\le& M_{h}&\le& 1\text{ TeV}\\
10\text{ GeV}&\le& M_{\nu'}&\le& 1\text{ TeV}\\
300\text{ GeV}&\le& M_{W'/Z'}& \le& 2\text{ TeV}\\
0&\le&g_h&\le&0.5\\
0&\le&g_Z&\le&0.5\\
0&\le&g_{Z'}&\le&0.5.
\label{eq:rhnranges}
\end{array}
\ee
where $M_{h}, M_{\nu'}$ and $M_{W'/Z'}$ are the Higgs boson, Dirac neutrino and $W'/Z'$ gauge boson masses, respectively.  The couplings of $\nu'$ to the Higgs, $Z$ and $Z'$ bosons are $g_{h},g_{Z}$ and $g_{Z'}$, respectively.  Since the Higgs boson couples the neutrino and matter  with Yukawa strength, it often has an inconsequential effect on the relic density and scattering rates.  The $Z$ and $Z'$ gauge bosons have gauge strength interactions, and the SM $Z$ boson has the strongest effect due to its lower mass.  In this scenario, where the $Z$ boson dominates in the calculation of both the relic density and elastic scattering cross section, the SI and SD cross sections are tightly correlated, as seen in Fig. \ref{fig:rhn}.  

This model predicts a rather large SI cross section that is partially excluded by limits from XENON10 and CDMS 2008.  The contours in Fig. \ref{fig:rhn} show the posterior distributions with the present SI and SD constraints enforced.  The predicted neutrino flux from annihilations in the Sun is significantly smaller than if the scattering limits were not included.  Nevertheless, the prospects for a significant signal in IceCube from this DM candidate are still good, as detailed in Section \ref{sect:icecube}.

\subsection{Minimal Universal Extra Dimensions}

Extra dimension models have gained popularity as a viable beyond the SM alternative~\cite{ArkaniHamed:1998rs,Randall:1999ee,Appelquist:2000nn,Hooper:2007qk,Rizzo:1999br,Kong:2005hn}.  Many types of extra dimensional models have been proposed.  Here we focus on the minimal Universal Extra Dimensions model (mUED)~\cite{Appelquist:2000nn}.  The stability of the DM candidate is ensured by KK-parity, which results from momentum conservation in the extra dimension. 

In the mUED model the DM has spin-1; it is the KK partner of hypercharge gauge boson, $B_1$.  The predictions of the KK particle spectra are mainly determined by one significant parameter, the curvature of the extra dimension, $R$.  Additionally, radiative corrections from the higher KK levels shift the masses of the KK states in the first level.  The number of KK levels included in the mass renormalization is given by $\Lambda R$, where $\Lambda$ is the cutoff scale of the model.  Many KK states from the first excitation in this model are nearly degenerate with $B_1$, requiring inclusion of coannihilation processes in the calculation of DM relic density.  After the relic density constraint is imposed, the curvature is found to be about $R^{-1}\sim 600-700$ GeV.    We allow the MCMC to vary over parameter ranges
\be
\begin{array}{ccccc}
114\text{ GeV}&\le& M_{h}&\le& 600\text{ GeV}\\
400\text{ GeV}&\le& R^{-1}&\le& 800\text{ GeV}\\
5&\le& \Lambda R& \le& 100.
\label{eq:uedranges}
\end{array}
\ee
where the upper bound on the Higgs mass~\cite{Gogoladze:2006br,Hooper:2007qk} is from electroweak precision constraints.

\begin{figure}[htbp]
\begin{center}
      \includegraphics[angle=0,width=0.47\textwidth]{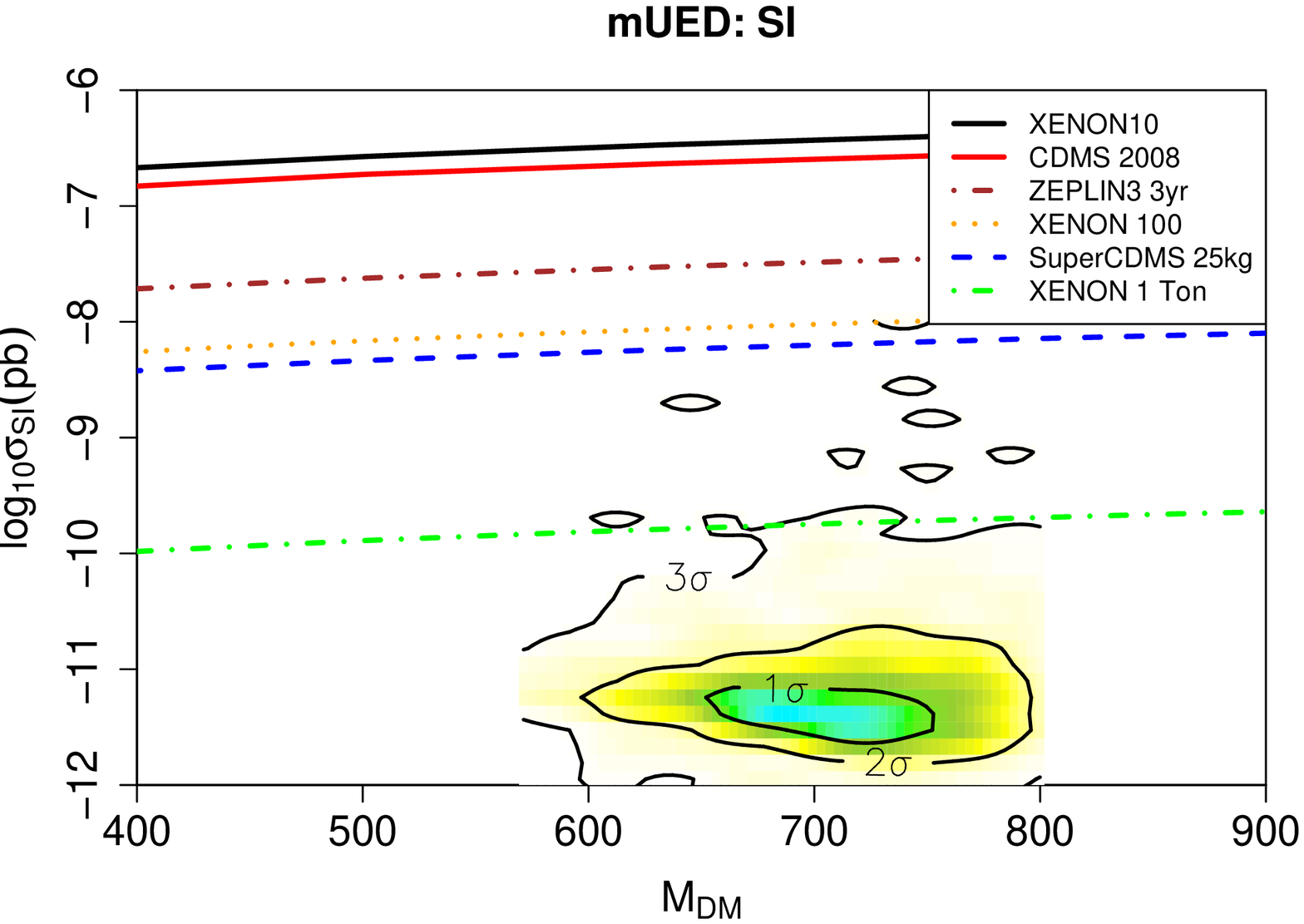}
      \includegraphics[angle=0,width=0.47\textwidth]{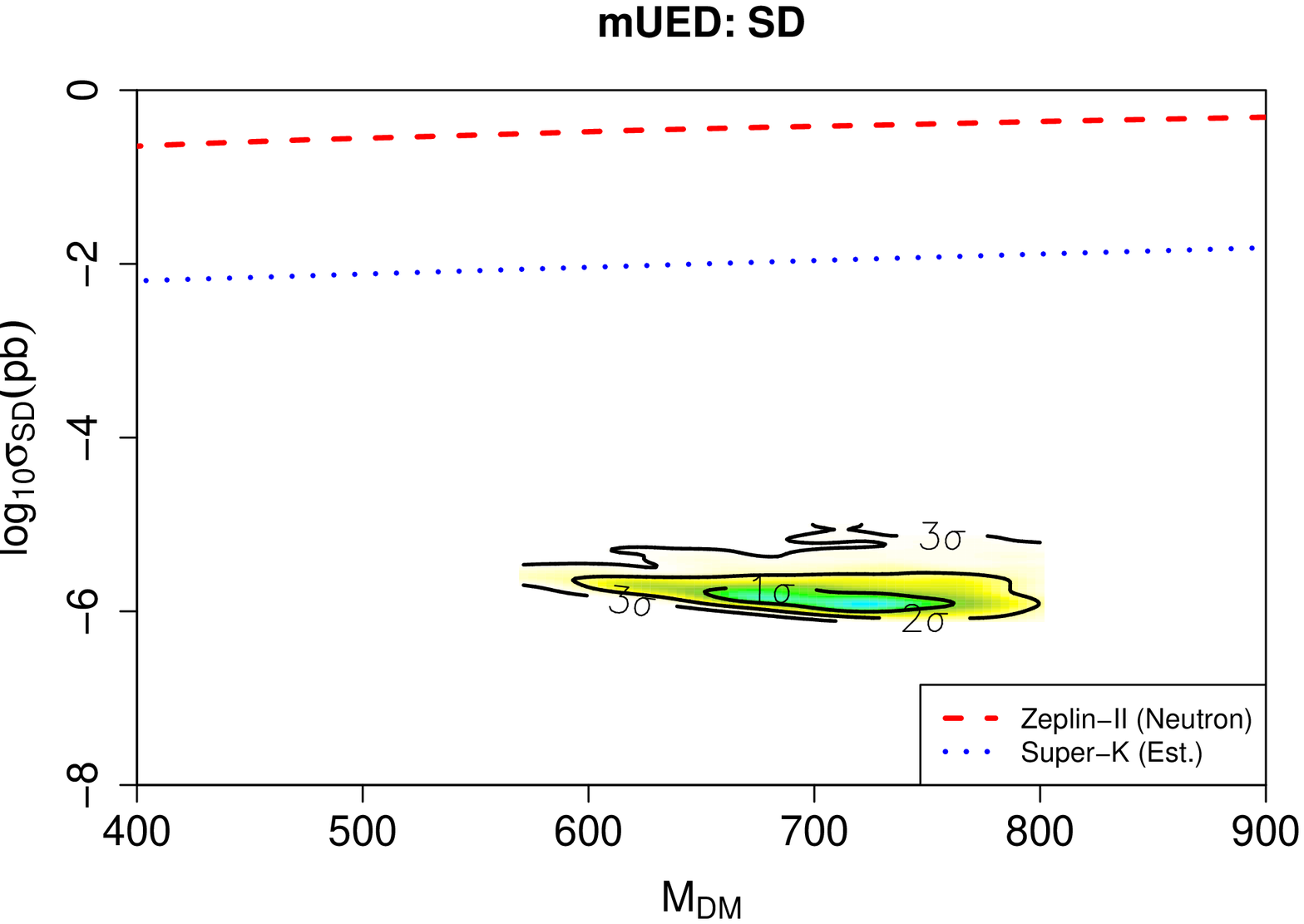}
      \includegraphics[angle=0,width=0.47\textwidth]{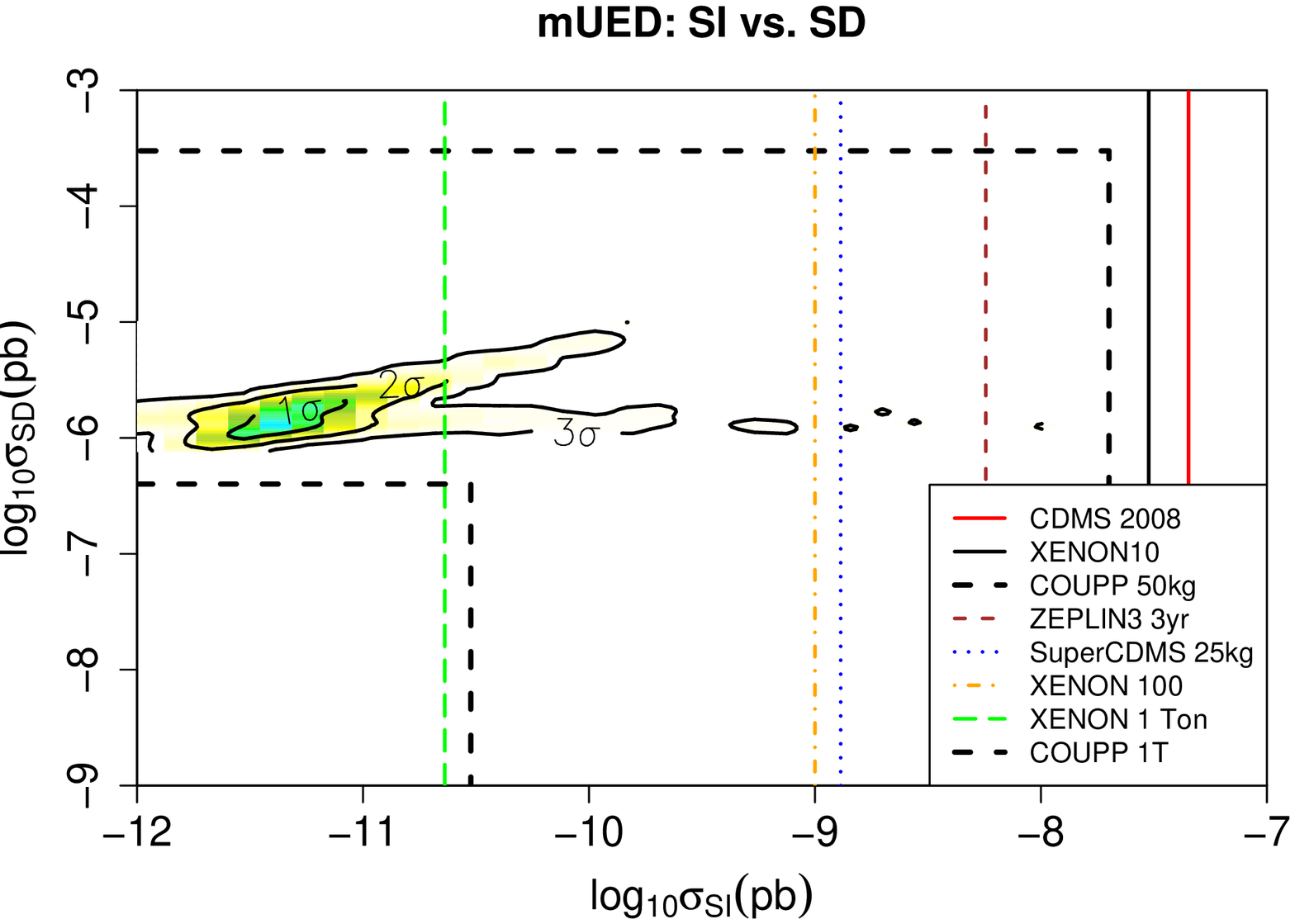}
\caption{mUED  predictions of the SI and SD cross sections.  The mUED predictions for the SD cross section may be probed with a one Ton detector if it has sufficient sensitivity at $M_{DM}\sim 700$ GeV.}
\label{fig:uedsipsdp}
\end{center}
\end{figure}

The DM scattering cross sections in mUED are displayed in Fig. \ref{fig:uedsipsdp}.  This model has a predicted 'sweet spot' of $\sigma_{SD} \sim {\cal O}(10^{-6})$ pb  for which the relic density is reproduced;  the $\sigma_{SI}$ range is considerably dispersed in comparison.  The source of this narrow range for the SD cross section is the limit from the relic density on the curvature of the extra dimension.  Since the KK quarks have approximately the same mass as the inverse curvature, the SD cross section is closely tied to the relic density.  The SI cross section is more dispersed due to the relatively large variation of the Higgs boson mass.  The test of the direct detection predictions of this model may require a 1 Ton detector.  However, an observation of a neutrino flux resulting form $B_{1}$ annihilations in the Sun is possible: see Section \ref{sect:icecube}.

\subsection{ Littlest Higgs with $T$-parity}

In Little Higgs models, the Higgs is an approximate Goldstone boson with its mass protected by approximate global symmetries~\cite{ArkaniHamed:2001nc,Han:2003wu,Perelstein:2005ka,Schmaltz:2005ky}.  Additional states are introduced to preserve this global symmetry and keep the Higgs from receiving quadratically divergent radiative corrections.  The Littlest Higgs model is one of the simplest and popular models~\cite{Low:2004xc,Hubisz:2004ft}.  It has a $SU(5)/SO(5)$ symmetry breaking pattern which gives two gauged copies of $SU(2)\times U(1)$ that are subsequently broken to $SU(2)_{L}\times U(1)_{Y}$ at the scale $f\sim 1$ TeV.  

While the Littlest Higgs model protects the Higgs boson mass, it does not satisfy  electroweak precision constraints.  Large corrections to electroweak observables occur at tree-level.  This may be alleviated by imposing a discrete $T$-parity that interchanges the two copies of $SU(2)\times U(1)$~\cite{Cheng:2004yc,Cheng:2003ju,Belyaev:2006jh}.  Once $T$-parity is introduced (the LHT model), electroweak corrections scale logarithmically with the cutoff scale, $\Lambda \sim 4 \pi f$.  We assume that the Wess-Zumino-Witten anomaly~\cite{Hill:2007zv} does not break $T$-parity to have a viable DM candidate~\cite{Barger:2007df}.  

The parameter ranges that we allow for the LHT model are 
\be
\begin{array}{ccccc}
114\text{ GeV}&\le& M_{h}&\le& 1\text{ TeV},\\
500\text{ GeV}&\le& f&\le& 2\text{ TeV},\\
\half&\le&\tan\alpha&\le& 2,\\
0&\le& \kappa,\kappa_{l}& \le& 4,
\label{eq:lhtranges}
\end{array}
\ee
where the $\kappa, \kappa_{l}$ couplings in the Lagrangian (see Eq. 2.14 of Ref.~\cite{Hubisz:2005tx}) determine the masses of the $T$-odd quarks and leptons masses, respectively.  The angle $\alpha$ defines the mixing between the $T$-even heavy quark and the SM top quark.   We require the breaking scale, $f$, to be below 2 TeV to avoid a reintroduction of the hierarchy problem.  In addition, we included the electroweak precision constraints on the oblique parameters given in Ref.~\cite{Hubisz:2005tx} and calculated the additional $\chi^{2}$ contributions associated with these parameters according to Ref.~\cite{Barger:2007im}.  The electroweak constraints in this model can allow a larger Higgs mass than in the SM, which will result in a suppression of the SI cross section.  To satisfy experimental bounds on four-fermi operators, the values of the $T$-odd masses must satisfy the bound~\cite{Hubisz:2005tx}
\be
{M_{T-odd}\over \text{ GeV}}\lesssim 4.8\times 10^{-3} (f/\text{GeV})^{2}
\ee 

\begin{figure}[htbp]
\begin{center}
      \includegraphics[angle=0,width=0.47\textwidth]{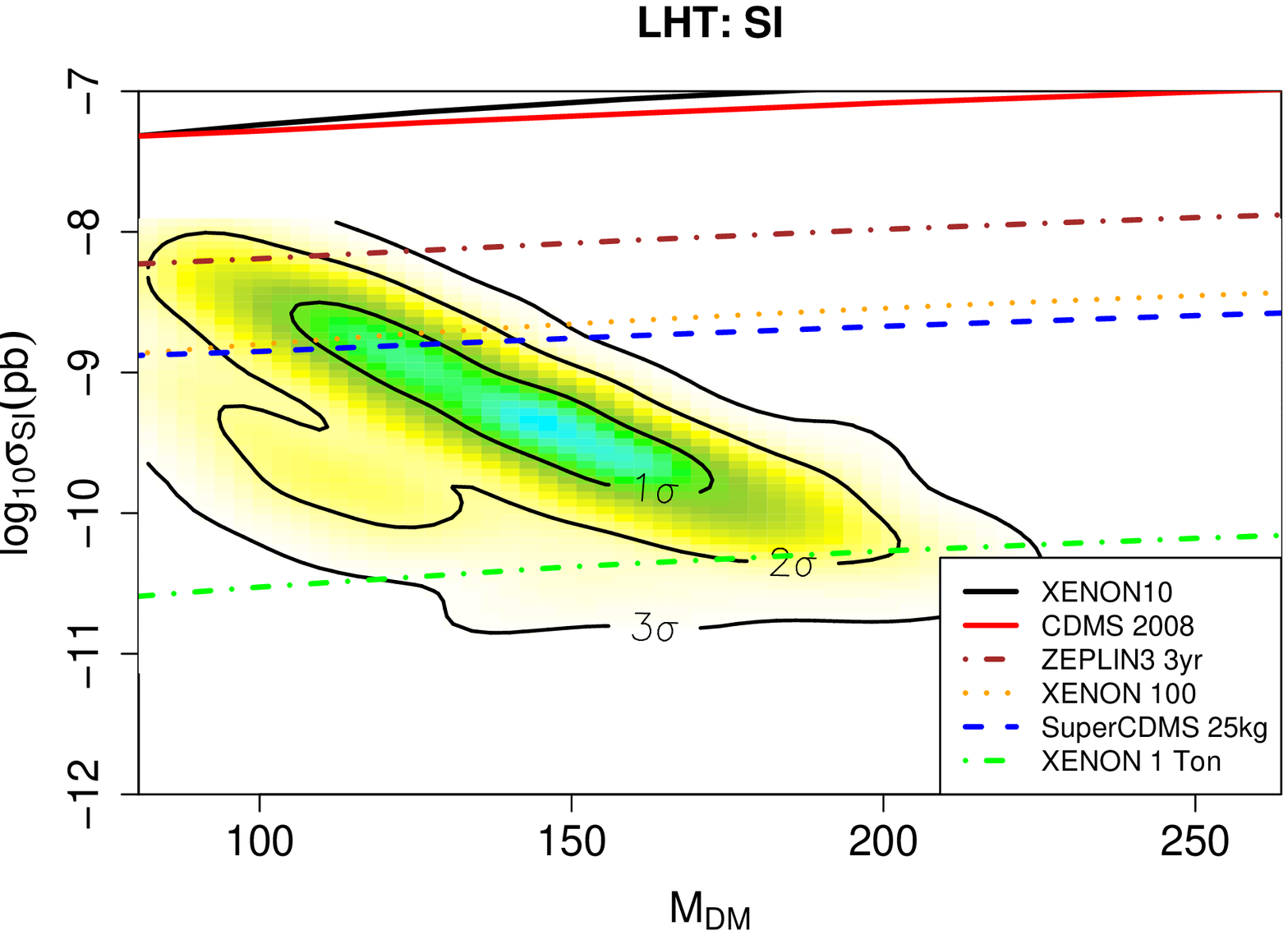}
      \includegraphics[angle=0,width=0.47\textwidth]{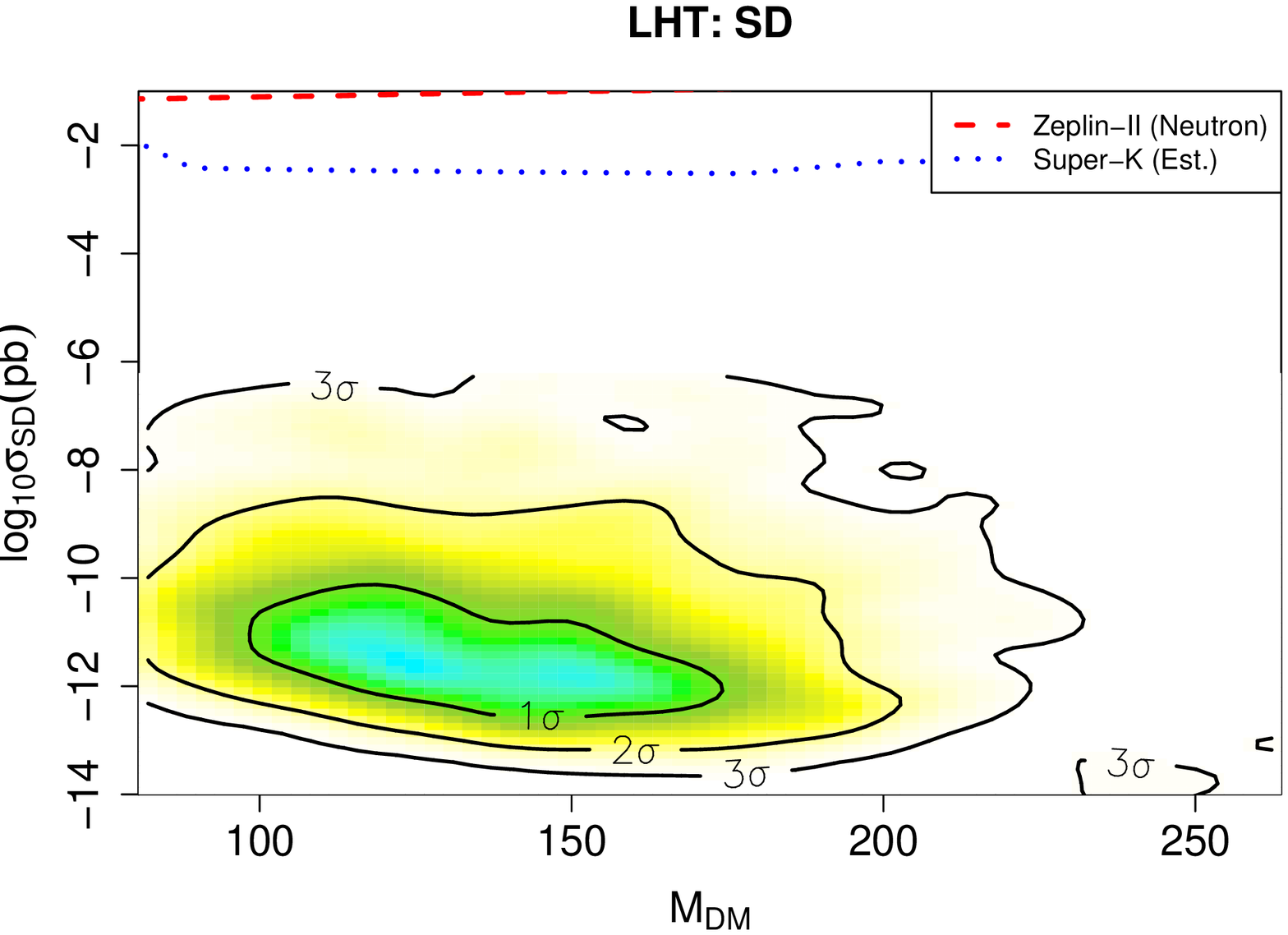}
      \includegraphics[angle=0,width=0.47\textwidth]{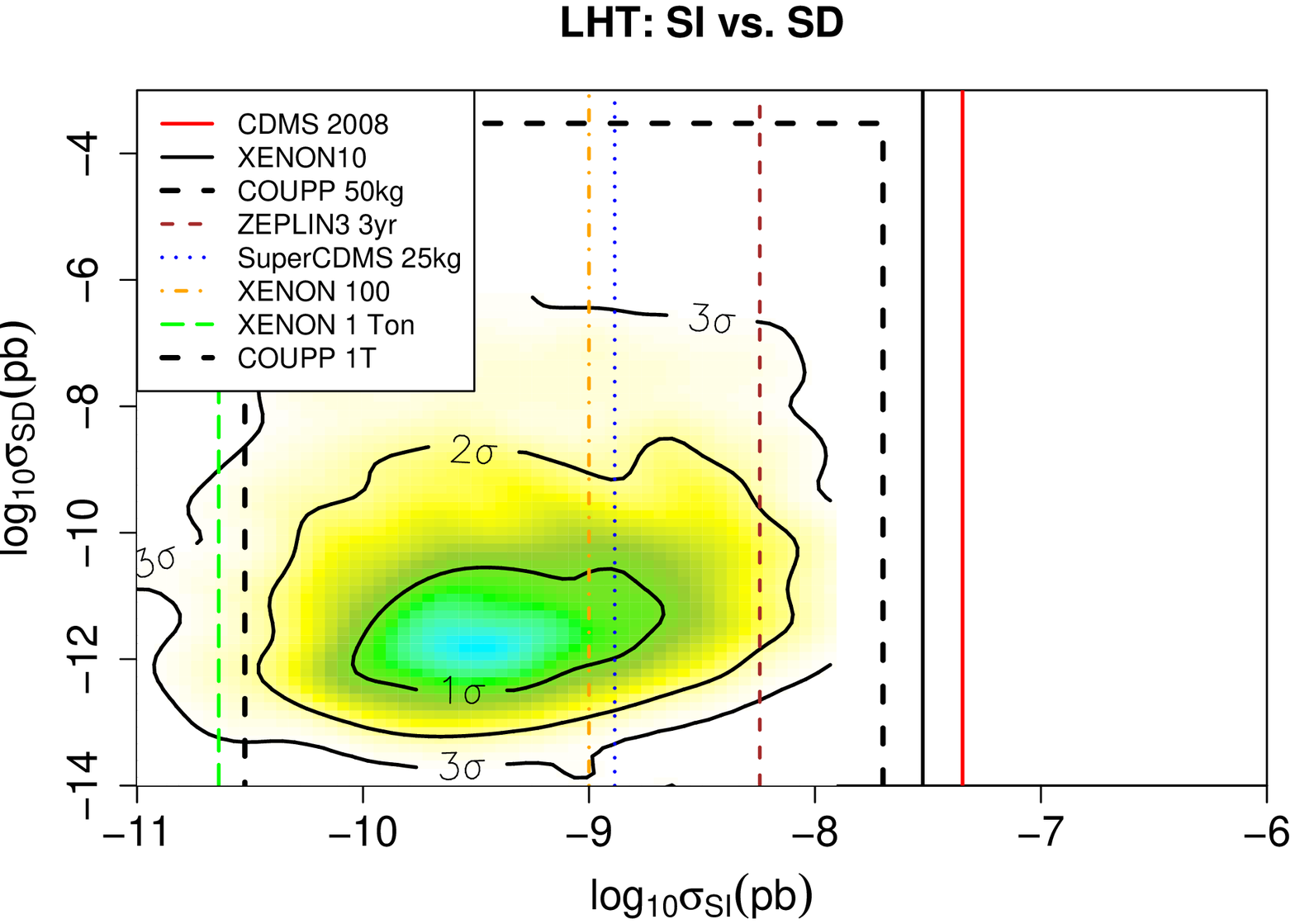}
\caption{Littlest Higgs model with $T$-parity predictions of the SI and SD cross sections.  The SD scattering is highly suppressed.}
\label{fig:lhtsipsdp}
\end{center}
\end{figure}

Associated with the introduction of  $T$-parity, there is a stable spin-1 DM candidate.  All non-SM gauge bosons and the triplet Higgs are $T$-odd, while SM fields are T-even.  The lightest $T$-odd state is the heavy photon, $A_{H}$.  The relic density is quite sensitive to the Higgs mass as the heavy photon annihilates dominantly through the Higgs boson.  Other annihilation processes involving the new heavy quarks are generally suppressed their large masses and the small hypercharge, $Y=1/10$ of the DM particle.  The SI scattering proceeds through the Higgs and $T$-odd quarks, giving SI cross sections of order ${\cal O}(10^{-9}-10^{-10})$ pb.  

We present the SI and SD cross sections in Fig. \ref{fig:lhtsipsdp}. Since SD scattering of the DM heavy photon occurs only via the $T$-odd quarks, the SD cross section is suppressed.
However, the predicted range of the SI cross section is within the projected sensitivities of future one Ton experiments.  A modification of the LHT model has been recently proposed that has the DM candidate couple more strongly to the Higgs boson~\cite{Bai:2008cf}.  The coupling enhancement  is about a factor of 10 larger than that of the standard LHT model and the model yields a much higher SI cross section.

\section{DM Scattering uncertainties}

\begin{figure}[htbp]
\begin{center}
      \includegraphics[angle=0,width=0.47\textwidth]{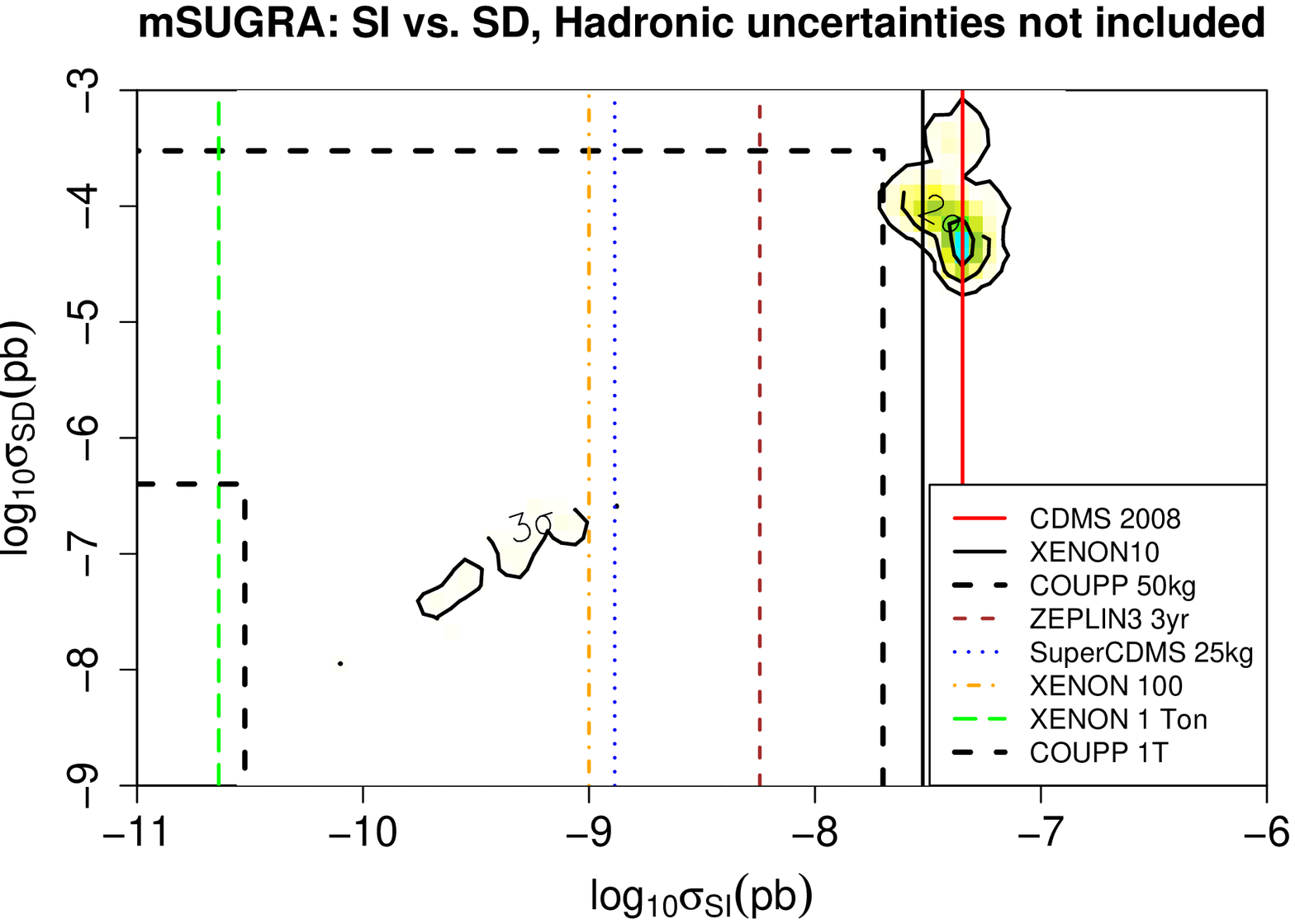}
      \includegraphics[angle=0,width=0.47\textwidth]{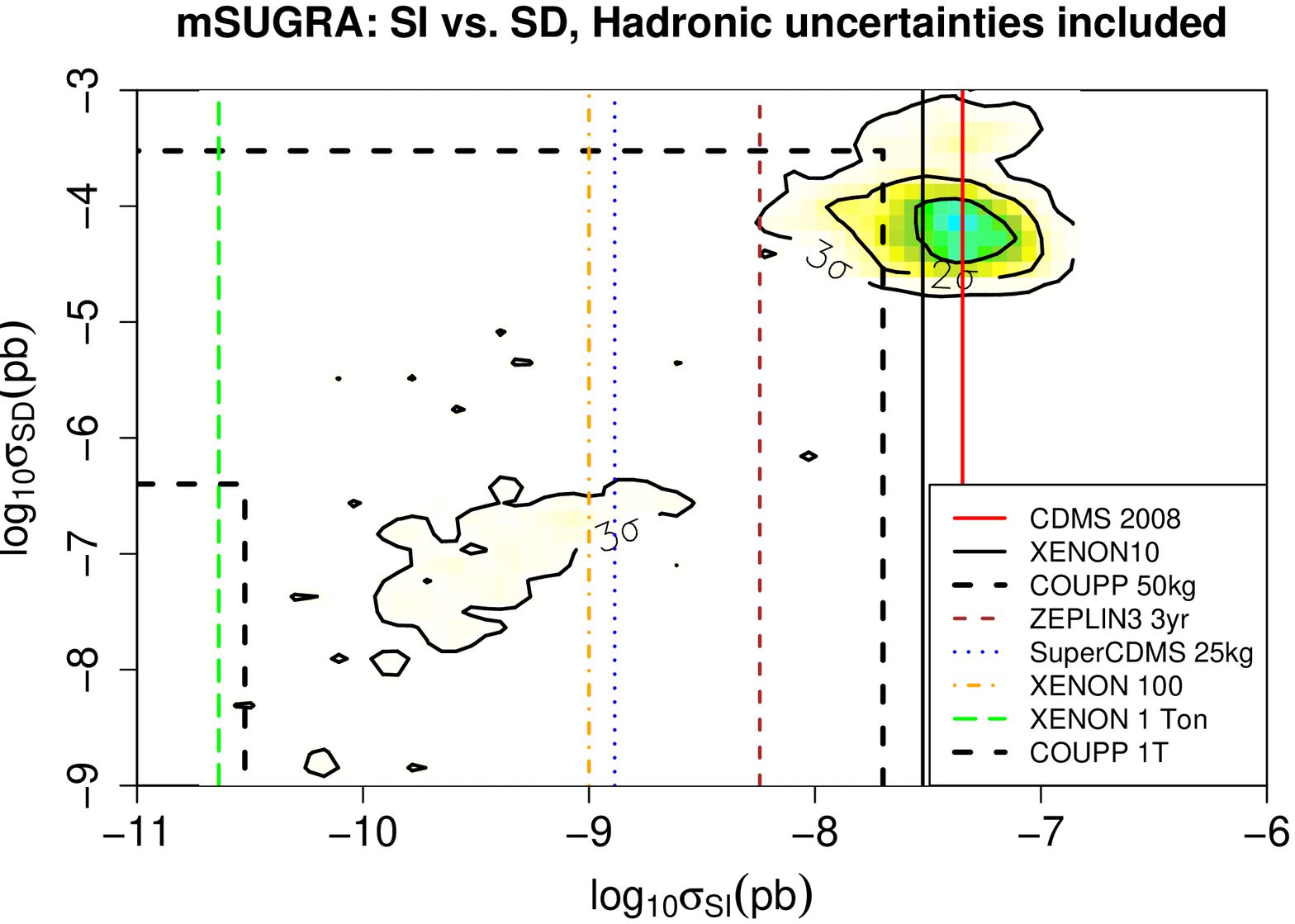}
\caption{Uncertainties in the sigma terms can produce substantial uncertainties in the SI and SD cross section predictions.  We illustrate this effect in the mSUGRA model; the left panel neglects these uncertainties  and the right panel includes the uncertainties.}
\label{fig:hadunc}
\end{center}
\end{figure}

Significant uncertainties on the hadronic matrix elements~\cite{Ellis:2005mb} are associated with the values of $\sigma_{\pi N}$ and $\sigma_{0}$: c.f. Section \ref{sect:predict}.  The dominant contribution to SI scattering is Higgs exchange coupled to the strange quark.   The uncertainty on the strange quark contribution causes variations in the SI cross section by an average of about $\pm30$\%.  In the left panel of Fig. \ref{fig:hadunc}, we present the posterior distributions of the SI and SD cross sections in the mSUGRA model with the uncertainties in the sigma terms neglected.  In the right panel of Fig. \ref{fig:hadunc} we include the uncertainties on the hadronic matrix elements.  These uncertainties substantially expand  the regions in the SI-SD plane encompassed by the predictions.  In all other illustrations in this paper we include the hadronic uncertainties.

\section{Expected signal rates at IceCube}
\label{sect:icecube}

High energy neutrinos are expected from the annihilations of DM that accumulated via gravitational capture in the Sun~\cite{Jungman:1994jr,Jungman:1995df,Barger:2001ur,Cirelli:2005gh}.  Neutrino telescopes are well positioned to search for neutrinos of this origin~\cite{Halzen:2005ar}.  The IceCube experiment at the South Pole is underway and expects to have about 50,000 events from atmospheric neutrinos in the near future~\cite{GonzalezGarcia:2005xw,Achterberg:2007bi}, demonstrating the capabilities of the detector to identify neutrino events.

\begin{figure}[htbp]
\begin{center}
      \includegraphics[angle=0,width=0.47\textwidth]{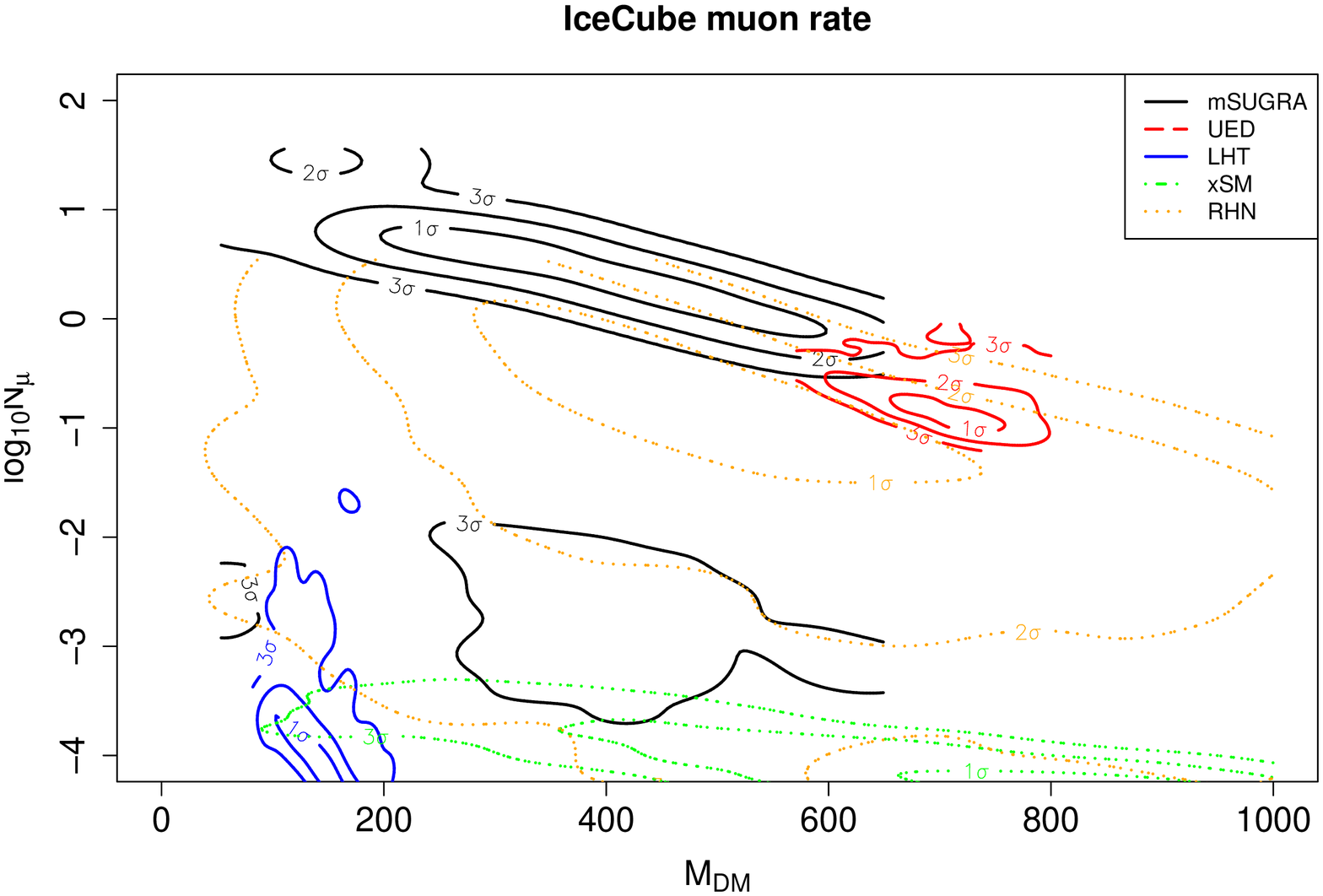}
      \includegraphics[angle=0,width=0.47\textwidth]{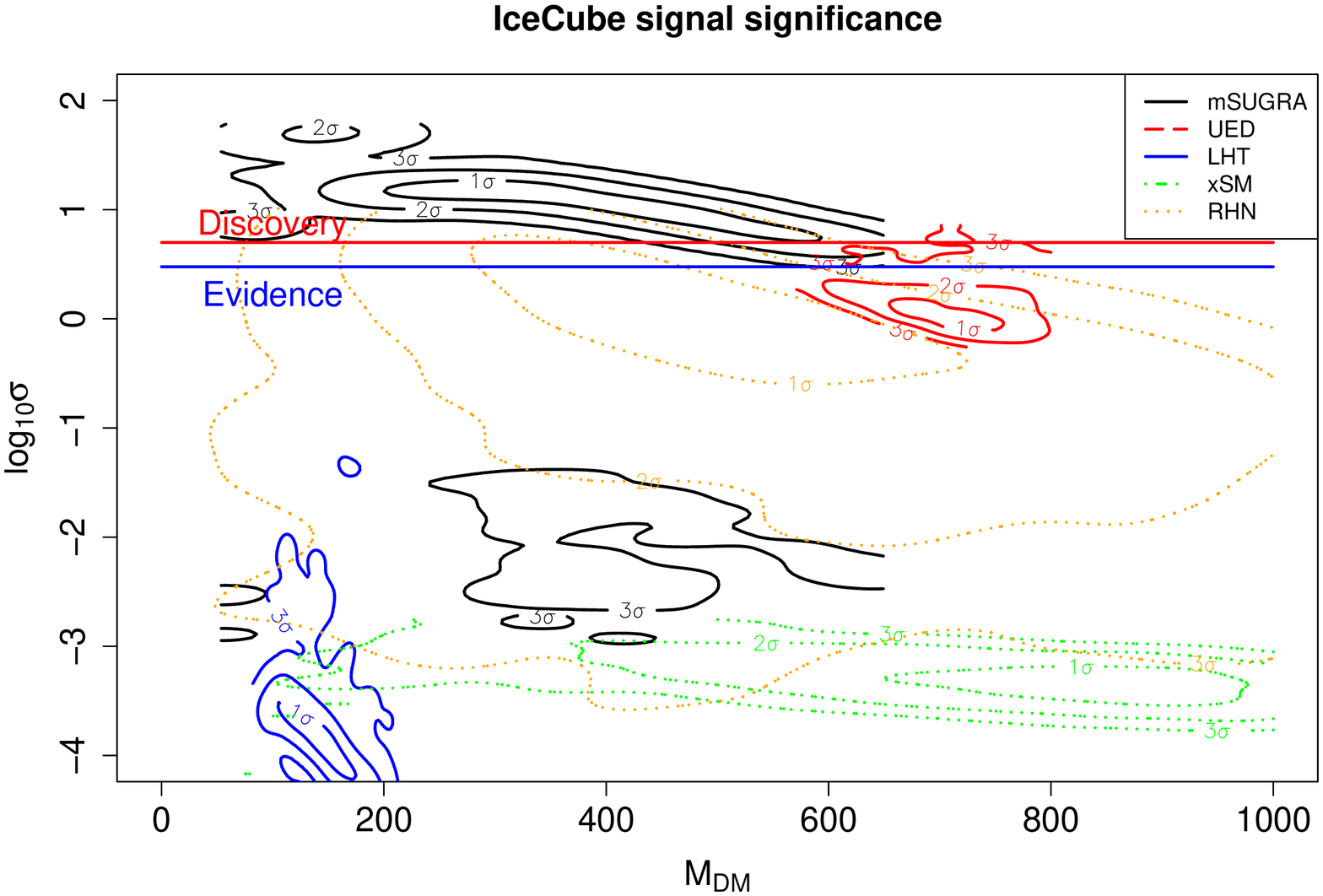}
\caption{The expected numbers of events and the statistical significance of DM signals for 3 and 5 years running of the IceCube neutrino telescope.  Only the FP region in mSUGRA and a portion of the mUED parameter space give a significance $>5\sigma$ for 5 years of running.  }
\label{fig:icecube}
\end{center}
\end{figure}

The basic requirements for a strong signal in km$^{2}$ sized neutrino telescopes from DM annihilations in the Sun are as follows:
\bi
\item A relatively large DM annihilation rate either directly to neutrinos or to SM particles that subsequently decay to high energy neutrinos (gauge bosons, top and bottom quarks, and tau leptons). The latter is generically the case for all the models considered.  
\item The DM mass exceeds the muon energy threshold of the neutrino telescope, which is $E_{\mu}\approx 50$ GeV for IceCube~\footnote{A future inner ring of photomultiplier strings in IceCube may lower this energy threshold to 10 GeV.}.  Since the DM annihilation proceeds non-relativistically, the DM mass is the maximum energy that any neutrino can have.  This kinematics essentially eliminates current IceCube tests of the tadpole nMSSM model since the lightest neutralino mass is typically   only 20-30 GeV, although in some small, fine-tuned portions of parameter space, the DM mass may extend above the $Z$-resonance.  
\item A sufficiently large SD cross section for neutralino captures (see Eq. \ref{eq:caprate}) to produce a detectable neutrino flux. A SD cross section $\gtrsim 10^{-4}$ pb is necessary to yield a neutrino flux that could be observed by IceCube.  Models that have either no or a vanishingly small SD cross section have no chance of producing an observable signal in km$^{2}$ sized neutrino telescopes.  This includes the xSM and the Littlest Higgs with $T$-parity.  
\ei

The predictions for IceCube of the models with SD scattering are given in Fig. \ref{fig:icecube}.  The numbers of muon events and the corresponding statistical significances for the models are given for both 3 and 5 years exposure with the full array (50\% of the array is now in place).  The calculations follow Ref.~\cite{Barger:2007xf}.  The FP region of mSUGRA bodes well for the detection of these neutrino events;  the predictions of the other models are below the $5\sigma$ discovery level; the RHN and mUED models are on the borderline of 3$\sigma$ detectability.

\section{Model Independent DM Scattering Cross Sections}
\label{sect:modindep}

In this section we categorize the interactions of DM candidates according to their intrinsic spins and provide generic expressions for the SI and SD cross sections.

\subsection{Self-Conjugate Spin-0}

A self-conjugate Spin-0 DM particle, $\phi$, cannot have a SD
scattering amplitude at threshold.  The generic effective Lagrangian
is
\begin{equation}
{\cal L}_\phi \supset
 {m_q\over \Lambda^2} \phi \phi \
\bar q (S_q + i P_q \gamma^5) q
\ .
\label{eq:spin0}
\end{equation}
The interaction, flipping  chirality, is proportional to the quark mass $m_q$.
Other forms of the interaction   can be reduced to this form by
the equation of motion\cite{Dobrescu:2007ec} with an emerging  mass
$m_q$ factor. 
Since the scalar DM candidate is self-conjugate, 
the vectorial part of the Lagrangian 
vanishes\footnote{This is in close analogy that the self-conjugate 
Majorana fermion cannot have a vectorial coupling.}.  
The pseudoscalar coupling, which is derivative in nature,
is not a leading contribution to the scattering because
the scattering amplitude is proportional to the DM velocity. The surviving scalar interaction
$S$ is purely SI.
For a nucleus target ($N$)  of $Z$ protons and $A-Z$ neutrons, 
the accumulated  amplitude is proportional to
the nucleus factor ${\cal C}_N$,
\begin{equation}
 {\cal C}_N = {2\over 27}   (S_c+S_b+S_t)A f^p_{TG}
+    \sum_{q=u,d,s} S_q(Z f^p_{Tq} + (A-Z) f^n_{Tq} )  \ .
\end{equation}
Here the first grouping is the gluonic contributions induced by
the heavy quarks $c,b,t$, and the second grouping is the contributions of the light quarks,
where  $f^p_{Tq}$  is the fraction
of the proton mass that the light quarks
$(u,d,s)$  represent,
$f^p_{Tq}m_p  \langle p| p\rangle \equiv \langle p| m_q\bar q q|p\rangle$
with $f_{Tg}\equiv1-\Sigma f_{Tq}$ 
as the remaining gluon fraction.
Their values for the proton and the neutron are~\cite{Ellis:2000ds}
\bea
f^p_{Tu}=0.020\pm 0.004&,&\quad f^p_{Td}=0.026\pm 0.005,\quad  f^p_{Ts}=0.118\pm 0.062 ,\\
 f^n_{Tu}=0.014\pm 0.003&,&\quad f^n_{Td}=0.036\pm 0.008
\eea
The cross section for the scattering of a scalar dark matter particle
on a nucleus  target is 
\begin{equation}
d\sigma^{\phi N}_{SI}
={(2!)^2\over 4\pi \Lambda^4 } \left( {m_\phi  m_N\over  m_\phi +m_N}\right)^2
\left( {m_p\over m_\phi}  {\cal C}_N \right)^2 
\ .\end{equation}
The formulas above (and also below) for the cross sections are
understood to be valid in the low momentum tranfer limit. A form
factor can be included for the finite size of nucleus.
For a proton target $m_N=m_p$, and $A=Z=1$.

If DM is spinless, there is no amplitude for SD scattering
and no amplitude for static  DM pair annihilation into a light fermion pair.
For this reason, this model predicts no
mono-energetic neutrinos~\cite{Barger:2007hj}
from annihilations of DM in the solar core. However secondary
neutrinos with energy distributions arising from DM  to $WW$, $ZZ$, $t\bar t$ etc. are still allowed, as studied in \cite{Barger:2007xf}.

\def\dum{
Since the cross sections from scattering from a proton and neutron are
expected to be nearly the same, we use the proton in our calculations.  Since
this cross section is constrained to be below the limits found by
XENON10~\cite{Angle:2007uj}, we show below that the prospects for
detection of scalar DM via the neutrino line are not promising.
The DM annihilation rate to neutrinos must be dominated by some scalar
exchange in the same manner as the DM scattering off matter above.
The possible candidates that can be exchanged include the Higgs boson
which couples through a loop to neutrino pairs or some unknown scalar
particle that couples to neutrinos.  In either case, the annihilation
rate to neutrino pairs is expected to be very weak.  This coupled with
the small scattering rate off matter imply that the chances for
discovering the DM candidate by its annihilations to neutrino pairs
using a neutrino telescope is very bleak.
}

\subsection{Dirac Fermion}
If the DM particle is a Dirac fermion, the generic 
interaction is described by
vectorial/axial-vector and scalar/pseuodoscalar couplings,
\begin{eqnarray}
 {\cal L}_\psi &\supset&  {1\over \Lambda^3}
\    \bar\psi (S_\psi+i\gamma^5 P_\psi    )\psi 
\    \bar q m_q (S_q+P_q i\gamma^5) q
\nonumber\\
& & +{1\over\Lambda^2}
\    \bar\psi \gamma_\mu (V+A\gamma^5)\psi
\    \bar q \gamma^\mu (v_q+a_q\gamma^5) q
\nonumber\end{eqnarray}
The vector-vector coupling $Vv_q$
and the scalar-scalar coupling $S_\psi S_q$  give rise to
the spin-independent (SI) low-energy scattering amplitude.
For a Dirac dark matter particle $\psi$, 
the SI cross section for scattering on a nucleus $N$ is
\begin{eqnarray}
 d\sigma_{SI}^{\psi N}= 
\left( {m^2_{\psi N}\over \pi \Lambda^4}\right) 
\left( {d\Omega\over4\pi} \right)
\left|   Vv_u(Z+A)+Vv_d(2A-Z)  +  S_\psi  (m_p/ \Lambda) {\cal C}_N\right|^2 
\ . \nonumber
\end{eqnarray}
In this formula we have used valence quark model relations.
Note that if the incident DM is an antiparticle $\bar\psi$, the cross
section is given by the same expression  except that the $V$ term flips sign.  Thus, particle $\psi$ and the
antiparticle $\bar\psi$ would scatter on a nucleus with different
cross sections when both vector-vector and scalar-scalar amplitudes
exist.

As SI measurements in $\psi+N$ scattering highly constrain 
the product $Vv_q$,  we shall neglect  $v_q$ for simplicity and focus our attention on 
the possibly much  larger SD scattering.
The vectorial part $V$ still plays a dominant role in the DM
annihilation but not in the SD $\psi+N$ scattering because its
time-component does not match the space-component of the quark axial
current.
The SD differential cross section can be written as
\begin{equation}
d \sigma_{SD}^{\psi+N} = {4\over\pi}
J(J+1) { m^2_{\psi N}\over  \Lambda^4}
\left| \sum_q  A a_q \lambda_q^{(N)}  \right|^2
\ {d\Omega\over 4\pi}\ 
\ ,\quad
m^2_{\psi N}= \left( {m_\psi m_N\over m_\psi +m_N} \right)^2
\ .
\end{equation}
where $J$ is the angular momentum of the target nucleus.
In case of the proton target that is relevant for solar capture of the DM,
$J={1\over2}$. The fractional quark-spin coefficient $\lambda_q^{(N)}$ is
defined in the Appendix.
All participating flavors $q$ are  summed at the amplitude level.
The reduced mass of $\psi+N$  appears as $m_{\psi N}$.  A direct search for DM is optimal with a target nucleus of
comparable mass to the DM particle and with  a large $J$ value.
A multiplicative unwritten form factor $F(Q^2)$ due to the nucleus size 
describes the dependence on the squared
momentum transfer, $Q^2$.  The $Q^{2}$ dependence may 
be significant for a large reduced mass.

If the DM couples to the $Z$ boson, the interaction can be written as
${\cal L} \supset \bar\psi \gamma_\mu (V+A\gamma_5)\psi Z^\mu$.
This effective amplitude can arise from the $t$-channel exchange
of $Z$ with the following relation
$$ {a_q\over \Lambda^{2}}=-{g\over 2\cos\theta_W M_Z^2} T_{3,q} \ ,\
   {v_q\over \Lambda^{2}}=
{g\over 2\cos\theta_W M_Z^2} (T_{3,q}-2Q_q\sin^2\theta_W) \ .$$
where $T_{3,u}={1\over2}$, and  $T_{3,d}=-{1\over2}$

Static annihilation of $\psi\bar\psi$
to a light fermion  pair can take place through the
vectorial amplitude $V$. The
 $\nu\bar\nu$ final state is of interest as a source of mono-energetic neutrinos\cite{Barger:2007hj} from the core of the Sun.  From DM annihilation through a $Z$ boson, or a new gauge boson of a larger gauge group under which DM
is charged,  the fraction of the DM annihilations to
neutrinos can, in principle, be quite large.  Models that allow the
possibility of Dirac DM include the hidden fermion in the Stueckelberg
$Z'$ model~\cite{Cheung:2007ut,Belanger:2007dx}.
Heavy sterile neutrinos also belong this type of model; they
constitute warm dark matter and are not considered in our study as
we assume the DM velocity is nonrelativistic~\cite{Dolgov:2000ew}.


\subsection{Majorana spin$-{1\over2}$ Fermion  }

For a spin $1\over2$ dark matter (DM) Majorana fermion,
the relevant dynamics is determined by the interaction
\be
 {\cal L} \supset 
 {1\over \Lambda^3}
\    \bar\chi (S_\chi+i\gamma^5 P_\chi    )\chi
\    \bar q m_q (S_q+P_q i\gamma^5) q
+
{1\over\Lambda^2}
  (\bar{\chi}   A \gamma_\mu\gamma_5\chi)
  \bar q \gamma^\mu (v_q+a_q \gamma_5) q  \ .
\ee
There is no vector-vector  interaction for the self-conjugate $\chi$ field.
The SI cross-section is given by the scalar-scalar interaction.
The $\sigma_{SI}$ formula is  like what is found in the Dirac case, 
with $V=0$ and an overall
factor of $(2!)^2$ due to two counts of Wick contraction for the
Majorana field,
\begin{eqnarray}
 d\sigma_{SI}^{\chi N}= 
\left((2!)^2  {m^2_{\chi N}\over \pi \Lambda^4}\right)
\left( {d\Omega\over4\pi} \right)
\ \left[   S_\chi  {m_p\over \Lambda} {\cal C}_N \right]^2
\ . 
\end{eqnarray}
The SD  scattering cross section is given by
\begin{equation}
d \sigma_{SD}^{\chi+N} = (2!)^2 {4\over\pi}
J(J+1) { m^2_{\chi N}\over  \Lambda^4}
\left| \sum_q  A a_q \lambda_q^{(N)}  \right|^2
\ {d\Omega\over 4\pi}\ 
\ .
\end{equation}
This result  agrees with Eq. (2.26) of
Ref. \cite{Ellis:1987sh},
provided that one is careful in distinguishing the amplitude and the
Wilson coefficient.


The annihilation of a pair of Majorana fermions $\chi\chi$ at rest
into a neutrino pair is forbidden because of
the helicity suppression from the extremely small mass of the neutrino.  The same helicity suppression applies to any light fermion pair.  The helicity suppression may be lifted, however, if a $Z$ boson is radiated
from an internal or external neutrino line~\cite{Barger:2006gw}.
However, this is a very weak process ${\cal O}(b_1^2\alpha^2
m_\chi^2/\Lambda^4)$ and it would also distort the neutrino line shape
considerably.  Models that allow this possibility include
SUSY~\cite{Baer:2006rs,Binetruy:2006ad,Drees:2004jm}, singlet femion DM~\cite{Babu:2007sm}, and
heavy Majorana neutrinos~\cite{Kitazawa:1996dp}.  For a detailed study
of neutrino signals from neutralino annihilation in SUSY through
intermediate states such as $W$ and $Z$ bosons and top quarks, see
Ref. \cite{Barger:2007xf}.

\subsection{Spin-1}
A spin-1 DM particle is encountered in Universal Extra Dimensions
(UED)~\cite{Antoniadis:1990ew,Randall:1999ee,Appelquist:2000nn,ArkaniHamed:1998nn,Nath:1999mw,Dicus:2000hm,Macesanu:2005jx,Kong:2005hn},  Littlest Higgs with $T$-parity (LHT)~\cite{ArkaniHamed:2002qy,Han:2003wu,Schmaltz:2005ky},
and Randall-Sundrum (RS) models~\cite{Randall:1999ee}, with an effective lagrangian
\begin{equation}
{\cal L} \supset 
{1 \over \Lambda^2}
     B^\mu B_\mu \ 
    S_q \bar  q m_q q   +
{A_q\over \Lambda^2}
i \epsilon^{\alpha\mu\nu\beta}
(B_\nu i{\stackrel{\leftrightarrow}{\partial}}{\ }_\alpha  B_\mu )
\ { \bar q }\gamma_\beta\gamma_5    q \ .
\end{equation}
The SI cross section has the same form as that of scalar DM,
\begin{equation}  d\sigma^{BN}_{SI}
={(2!)^2\over 4\pi \Lambda^4} \left( {m_N m_B\over  m_N+m_B}\right)^2
\left( {m_p\over m_B}  {\cal C}_N \right)^2  \ . \end{equation}
The SD scattering amplitude
$B(\epsilon_i)+N \to B(\epsilon_f)+N $ is given in terms
of the incoming (outgoing) polarizations
$\epsilon_i$ ($\epsilon_f$) by the expression
\be
 {\cal M} = 2! {A_q\over \Lambda^2}
(\epsilon_f^*)_\mu
i\epsilon^{0\mu\nu\beta}
(\epsilon_i)_\nu    (2 m_B)
\ \langle N | { \bar q}\gamma_\beta\gamma_5    q |N\rangle
\ 2m_N
\ee
where the factor $2m_N$ links the nonrelativistic
convention $ \langle N|\bar q \BM{\gamma}\gamma_5 q|N\rangle
= 2\lambda^{(N)}_q  \langle N|{\bf J}_N |N\rangle$
to the relativistic normalization.
The combinatorial factor 2! is due to the self-conjugate $B_1$ field in
the Lagrangian operator.  The amplitude simplifies to
\be
 {\cal M} ={2! \over \Lambda^2}
A_q (2 m_B) \ 2m_N
(\epsilon_{\rm f})^*_\ell
i\epsilon^{\ell k j}
(\epsilon_{\rm i})_j
\cdot  \langle N | { \bar q} \gamma_k \gamma_5    q |N\rangle
\ee
The spin-1 matrix is $(S_k)_{\ell j} =i \epsilon_{\ell k j}$.
Therefore, the amplitude takes the form
\be
 {\cal M} ={2! \over \Lambda^2}
(2 m_B) \ (2m_N)\ A_q   2\lambda_q^{(N)} \BM{S}\cdot \BM{J}  \ . 
\ee
The SD scattering cross section is given by
\begin{equation}
d \sigma_{SD}^{B+N} = (2!)^2 {8\over3\pi}
J(J+1) { m^2_{B N}\over  \Lambda^4}
\left| \sum_q  A_q \lambda_q^{(N)}  \right|^2
\ {d\Omega\over 4\pi}\  
\ .
\end{equation}
Like fermionic DM,
the SD operator does not give a static spin-1 DM annihilation
amplitude  to a fermion pair. However, there are other effective
operators, as outlined  in the Appendix, that are relevant to fermion
final states.  Mono-energetic neutrinos can be produced from spin-1 DM
annihilation in the solar core.

\subsection{Spin $3\over2$}
For the DM particle that is described by a self-conjugate
Rarita-Schwinger field $\Psi_\mu$ of spin $S={3\over2}$,
like the gravitino $\widetilde G^0$ in supergravity (SUGRA),
the relevant  interaction can be written as
\be {\cal L} \supset  
{1\over \Lambda^3} \bar\Psi_\mu S_\Psi \Psi^\mu 
\ \bar q S_q m_q q       +
{A_q \over \Lambda^2}  {3\over2}
(\bar \Psi_\mu  \gamma^\beta\gamma_5\Psi^\mu )
(\bar q \gamma_\beta\gamma_5 q)   \ .  
\ee
The SI formulas are like those of the spin-${1\over2}$
Majorana case.
\begin{eqnarray}
 d\sigma_{SI}^{\Psi N}= 
\left((2!)^2  {m^2_{\Psi N}\over \pi \Lambda^4}\right)
\left( {d\Omega\over4\pi} \right)
\ \left[   S_\Psi  {m_p\over \Lambda} {\cal C}_N \right]^2
\ . \nonumber
\end{eqnarray}
The SD amplitude in our special case $S=3/2$ turns out to have a universal form,
\be {\cal M}(\Psi+N\to\Psi+N) =  {2!\over\Lambda} A_q \ (2m_\Psi)\ (2m_N)
\BM{S} \cdot \ (2\lambda_q^{(N)}  \BM{J})   \ .\ee
The square is simplified with the use of the identity
Tr$[S^a   S^b]=\delta^{ab}(2S+1)S(S+1)/3$ when the spin orientation of
$S$ is summed. The spin-averaged square of the elastic SD scattering amplitude is
\be {\sum}_{m_S} |{\cal M}|^2 =
\left|\sum_q  {A_q\over \Lambda^2}  \lambda_q^{(N)} \right|  ^2
\ \hbox{$64\over3$} M^2 m_N^2  J(J+1) (2S+1)S(S+1)  \ee
Averaging over the $(2S+1)$ spin components, we obtain
\be d \sigma_{SD}(\Psi N)={16\over3\pi} S(S+1) J(J+1)
{m_{\Psi N}^2\over\Lambda^4}
\left| \sum_q A_q \lambda_q^{(N)} \right|^2
\ {d\Omega\over 4\pi}\ |F(Q^2)|^2 \ee
With  appropriate substitutions,
this generic formula agrees with that   of
the self-conjugate spin-1 case, as well as that of the
Majorana  spin-${1\over2}$ fermion, if we set $ A a_q ={1\over2}A_q $.

\section{Conclusions}
\label{sect:concl}
Both direct detection experiments for DM scattering and LHC experiments can discover a weakly interacting DM particle and reveal its properties.  In Section \ref{sect:predict} we showed that the SD versus SI cross section provides an excellent DM diagnostic.  We highlighted this by scanning over the parameter space of six Beyond the SM models which supply viable DM candidates:  mSUGRA, singlet extended SM and MSSM, a stable Dirac neutrino, Littlest Higgs with $T$-parity and Minimal Universal Extra Dimensions.  To thoroughly scan the parameter spaces, we adopted the Bayesian method that is the foundation of the Markov Chain Monte Carlo approach.  The DM models can have distinct phenomenological predictions.  We showed that  the DM model possibilities can be narrowed by measurements of both SI and SD elastic scattering.  The direct signals for DM in recoil and neutrino telescope experiments are complementary to LHC experiments in distinguishing the beyond the standard model physics scenarios~\cite{Hubisz:2008gg}.

\begin{figure}[htbp]
\begin{center}
      \includegraphics[angle=0,width=0.67\textwidth]{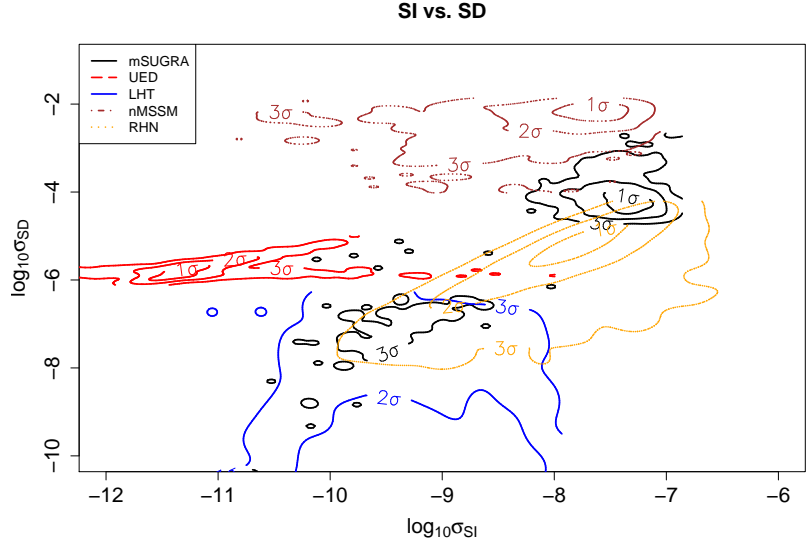}      
\caption{The posterior distributions of $\sigma_{SI}^{DM-p}$ and $\sigma_{SD}^{DM-p}$ for six models.}
\label{fig:allxsect}
\end{center}
\end{figure}

We  summarize below the model predictions for the DM cross sections; the posterior distributions are summarized in Fig. \ref{fig:allxsect}.  
\bi
\item In mSUGRA, the FP region provides the largest SI and SD cross sections.  This is due to the mixed Higgsino nature of the lightest neutralino; the neutralino couplings to the Higgs and $Z$ bosons are large.  The Bino nature of the lightest neutralino in the CA and AF regions causes these scenarios to have substantially smaller cross sections. 
\item The tadpole nMSSM model has large SD scattering, of order $10^{-3}$ pb,  and a wide range of SI cross section.  This is a consequence of the DM annihilation occuring through the $Z$ boson . To counter the small annihilation rate in the early universe (due to the small neutralino mass in the model), the neutralino pair is required to have a larger $Z$ boson coupling, resulting in a large SD cross section.  
\item In the singlet extended SM, the DM candidate is spin-0, which gives a vanishing SD cross section.  The SI cross section is generally small, below $\sim 10^{-8}$ pb, and occurs through Higgs boson exchange.  If SD scattering of DM is observed, the class of models with spin-0 DM would be immediately excluded as being the sole origin of the DM in the universe.  
\item For stable Dirac neutrino DM,  the $Z$ boson dominates in the calculation of both the relic density and elastic scattering cross section and makes the SI and SD cross sections tightly correlated and large.
\item In mUED, a 'sweet spot' of $\sigma_{SD} \sim {\cal O}(10^{-6})$ pb exists for which the DM relic density is reproduced.  The relic density is strongly dependent on the curvature parameter and fixes its value.  The KK quarks have approximately the same mass as the inverse curvature and the SD cross section is thus closely tied to the relic density.  On the other hand, the $\sigma_{SI}$ cross section is more dispersed due to the larger variation of the Higgs boson mass. 
\item In the LHT model, the SD interaction occurs through T-odd quarks which have a small hypercharge.  Therefore, the SD cross section in this model is typically very small.  In contrast, the SI scattering proceeds through the Higgs and $T$-odd quarks, giving experimentally accessible SI cross section values.

\ei

\begin{figure}[htbp]
\begin{center}
      \includegraphics[angle=0,width=0.47\textwidth]{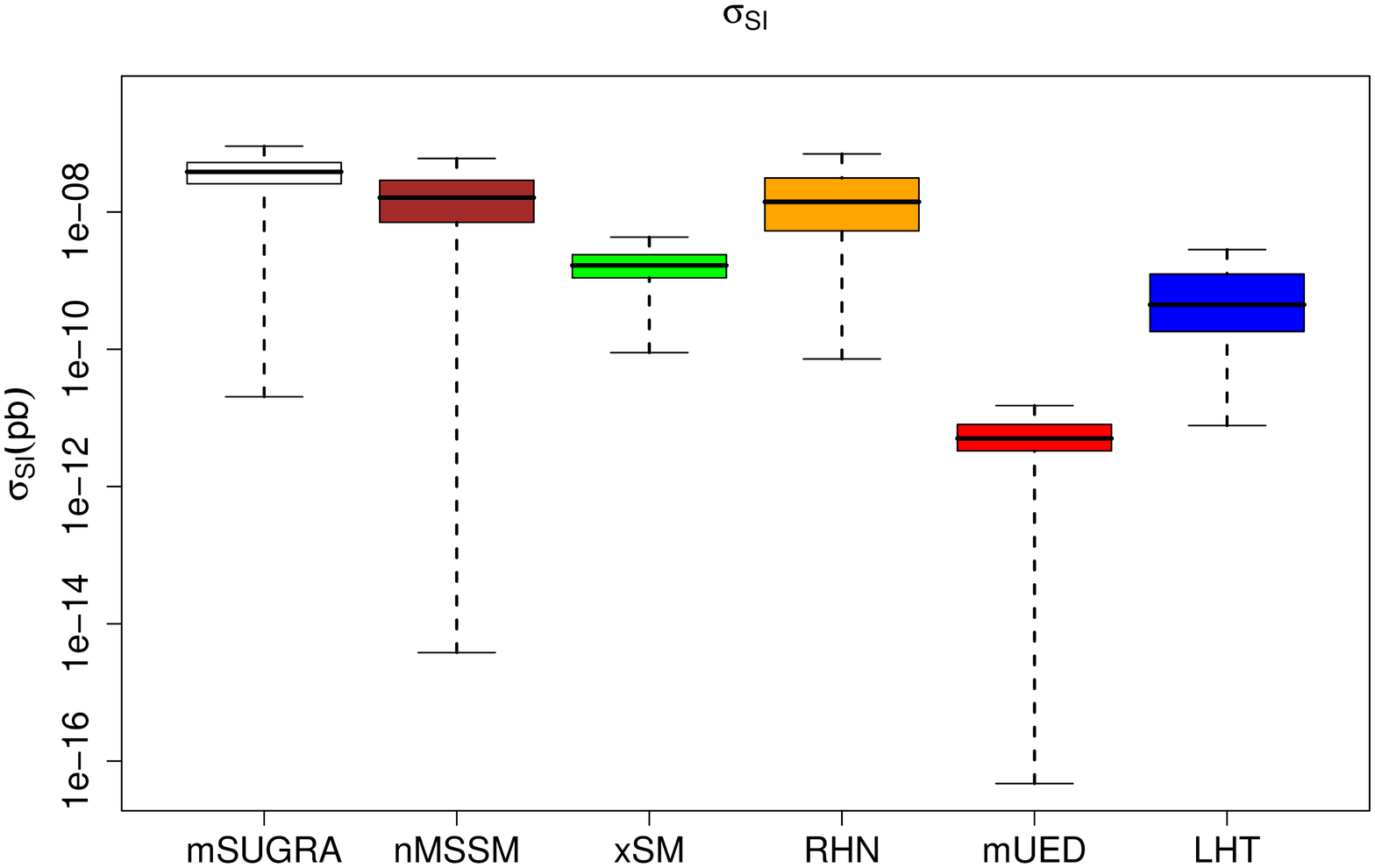}
      \includegraphics[angle=0,width=0.47\textwidth]{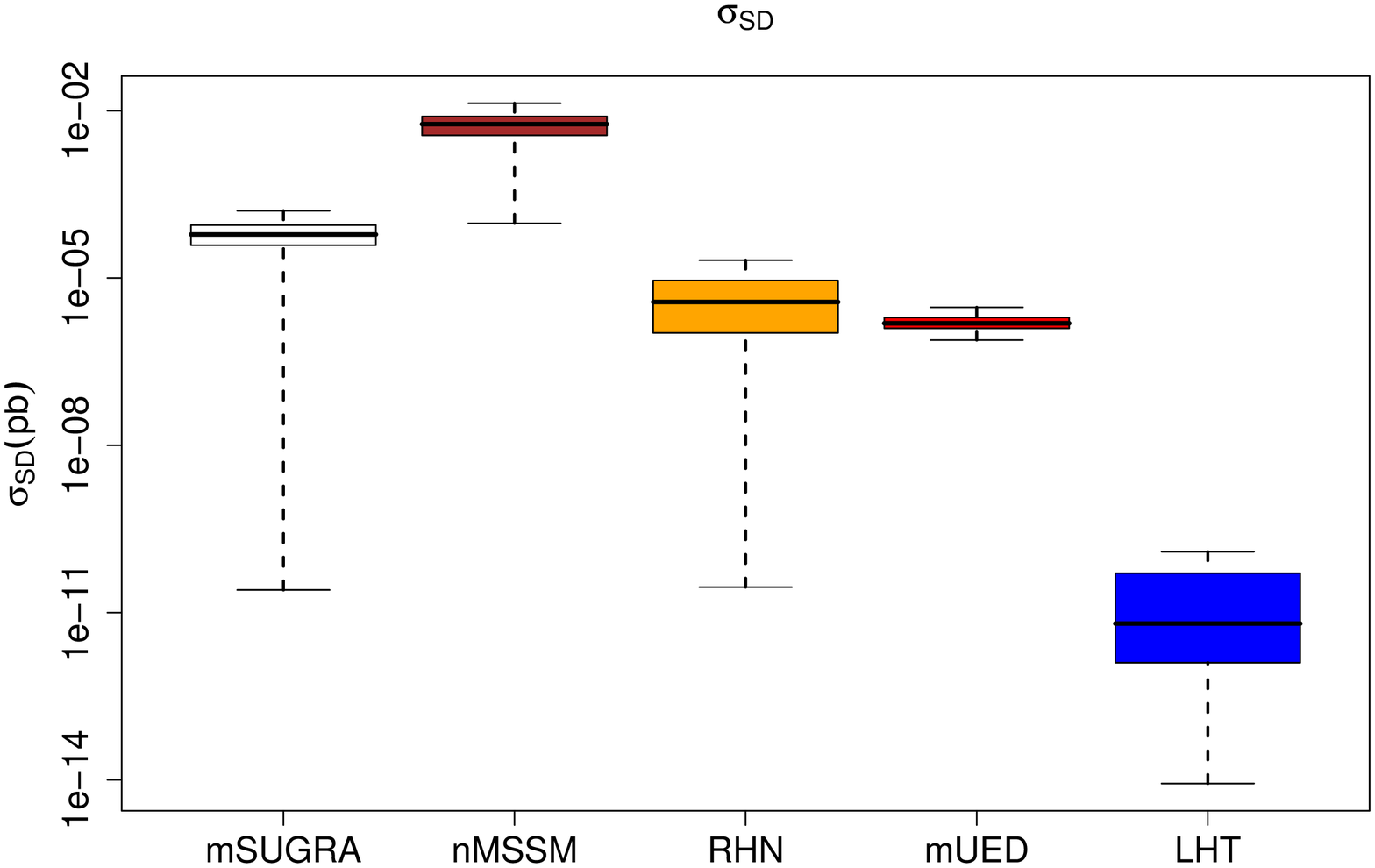}
\caption{Box and whisker plot of the SI and SD cross sections.  The box represents the coverage of the middle 50-percentile and the whisker is $\sim 100\%$ coverage.  The xSM is not shown in the SD panel as its cross section vanishes. The large SD cross section of the nMSSM is difficult to probe because of the low ($\sim 20-30$ GeV) mass of the DM particle in this model.}
\label{fig:boxwhisker}
\end{center}
\end{figure}
We provide a summary of the SI and SD cross sections by the box and whisker plots in Fig. \ref{fig:boxwhisker}.   The boxes represents the coverage of the middle 50-percentile.  
We summarize the forecast for observing a signal in neutrino telescopes or the scattering cross sections at future recoil experiments in Table \ref{tab:illust}.   

We have also presented model-independent parameterizations of the SI and SD cross sections for DM of arbitrary spins.

\begin{table}[htdp]
\caption{Survey of models and DM scattering predictions organized by the intrinsic spin of the DM candidate.  The forecast for IceCube detection is also given, wherein we describe in what portion of the parameter space a $5\sigma$ discovery can be made.  Instances where the likelihood of detection via IceCube are low are marked by $\times$.  The  $2\sigma$ ranges for $\sigma_{SI}^{DM-p}$ and $\sigma_{SD}^{DM-p}$ are given in pb units.}
\begin{center}
\begin{tabular}{|c|c|cccc|}
\hline
$S$ & Model & Candidate & IceCube Prospects & $\sigma_{SI}^{DM-p}$
&$\sigma_{SD}^{DM-p} $\\
\hline
0&xSM& $S$ & $\times$  &  $6\times10^{-10}-4\times10^{-9}$ & $\times$
\\
\hline
$\half$&mSUGRA& $\widetilde \chi_1^0$ & Majority of FP region &   $4\times10^{-10}-8\times10^{-8}$&   $1\times10^{-7}-2\times10^{-4}$ \\ 
$\half$&nMSSM& $\widetilde \chi_1^0$ & $\times$ by $E_{\mu}$ threshold &  $1\times10^{-9}-5\times10^{-8}$ &  $10^{-3}-10^{-2}$   \\
$\half$&RHN& $\nu'$ & Portion  & $9\times10^{-10}-9\times10^{-8}$  &  $7\times10^{-8}-3\times10^{-5}$ \\
\hline
$1$&mUED& $B_{1}$ & Portion &  $2\times10^{-12}-4\times10^{-11}$&  $1\times10^{-6}-3\times10^{-6}$ \\
$1$&LHT& $A_{H}$ & $\times$ by $\sigma_{SD}$ suppression &  $6\times10^{-11}-4\times10^{-9}$ &  $2\times10^{-13}-4\times10^{-9}$ \\
\hline
\end{tabular}
\end{center}
\label{tab:illust}
\end{table}%


\vskip1cm
\hrule
\vskip1cm

\begin{acknowledgments}
\end{acknowledgments}
We thank H. Baer, D. Cline, F. Halzen, A. McDonald, A. Noble, L. Roszkowski and A. Tregre for valuable discussions.  VB thanks the KITP Santa Barbara for hospitality.  This work was supported in part by the U.S.~Department of Energy under grant Nos. DE-FG02-95ER40896 and DE-FG02-84ER40173, the NSF under Grant No. PHY05-51164 and by the Wisconsin Alumni Research Foundation.

\appendix
\section{Nuclear Matrix and Form Factor}
The spin dependent (SD) cross section  involves
the matrix element of spatial components of
the quark axial-current $\bar q\BM{\gamma}\gamma_5 q$.
The matrix elements of the proton $(p)$ and the neutron $(n)$ are
\be \langle p|\bar q \BM{\gamma}\gamma_5 q|p\rangle
=2\Delta_q^{(p)}\ \langle p|{\bf s}_p |p\rangle
\ ,\
   \langle n|\bar q \BM{\gamma}\gamma_5 q|n\rangle
=2\Delta_q^{(n)}\ \langle n|{\bf s}_n|n\rangle
\ ,\ee
where $\Delta_q$  measures the
fraction of the  spin ($\BM{s}_q={1\over2}\bar q\BM{\gamma}\gamma_5 q$)
carried by the quark $q$ in the nucleon
as in \cite{Ellis:1987cy}.
Isospin symmetry relates $\Delta_q^{(p)}$ and $\Delta_q^{(n)}$,
{\it i.e.} $\Delta_u^{(p)}=\Delta_d^{(n)}$ etc.
From the
Wigner-Eckardt theorem,
the matrix element for a nucleus ($N$) has the
structure\cite{Goodman:1984dc}
\be \langle N|\bar q \BM{\gamma}\gamma_5 q|N\rangle
= 2\lambda^{(N)}_q  \langle N|{\bf J}_N |N\rangle  \ .\ee
For an isolated  single proton or neutron, the coefficients $\lambda_{a}$ are simply
\be \lambda_q^{(p)}= \Delta_q^{(p)} \ ,\
  \lambda_q^{(n)}= \Delta_q^{(n)}  \ .\ee
The numerical values\cite{Mallot:1999qb}
for a proton are
\be  \Delta_u^{(p)}=0.78 \pm 0.04 \ ,\  \Delta_d^{(p)}=-0.48\pm 0.04 \ ,\
   \Delta_s^{(p)}=-0.15 \pm 0.04 \ . \ee
The Wigner-Eckardt coefficient $\lambda^{(N)}_q$ relates
the quark spin matrix element to angular momentum of the nucleons.
For the case of a  nucleus with a single unpaired valence-nucleon ($p,n$),
the shell model gives the Land\'e formula,
\be\lambda_q = {1\over2}\{1+[s(s+1)-\ell(\ell+1)]/[J(J+1)]\}
\Delta_q^{(p,n)}  \
. \ee
For general cases the 
estimates of the coefficients are parametrized as
\be \lambda_q^N \simeq (1/J)[
     \Delta_q^{(p)}   \langle S_p\rangle
 +   \Delta_q^{(n)}    \langle S_n\rangle ]   \ ,
\ee
as in the review of Ref.~\cite{Jungman:1995df}.
The quantities $\langle S_p\rangle$  and $\langle S_n\rangle$  
represent the respective expectation values for the proton and neutron group spin
contents in the nucleus. They are very model dependent. Ref.\cite{Dimitrov:1994gc} evaluated
$\langle S_p\rangle=0.03$,
$\langle S_n\rangle=0.378$ for  $^{73}$Ge.
There is  also a multiplicative form factor
correction\cite{Dimitrov:1994gc}
$F(Q^2)$ to the scattering amplitude,
that can be straightforwardly incorporated.  The momentum transfer squared is
\be Q^2=|\BM{p}_N-\BM{p'}_N|^2 = m_r^2 v_r^2 \sin^2{\theta\over2} \ ,\ee
with $m_r$ the reduced mass, $v_r$ the relative velocity
of the DM particle and the target nucleus and $\theta$ the scattering angle.
It is expected that the form factor effects are only relevant for a
sizable nucleus.
Usually, a strong isospin decomposition is assumed in the parameterization. The isoscalar and  isovector parts are
\be a_0=(A_u+A_d)(\lambda_u+\lambda_d)+2A_s\lambda_s \ ,\quad
  a_1=(A_u-A_d)(\lambda_u-\lambda_d) \ . \ee
\be S(Q^2)=a_0^2 S_{00}(Q^2)+a_1^2 S_{11}(Q^2)+a_0a_1S_{01}(Q^2) \ee
Those $S$ functions are well documented in nuclear model analyses.
The form factor squared is given by 
\be
|F(Q^2)|^2 =S(Q^2)/S(0) 
\ee

\section{Spin-1 amplitude of $B(k,\mu) + f(p)  \to  B(\ell ,\nu) f(q) $}

We use the mUED model to illustrate the structure of the scattering amplitude.
The LHT  model follows a similar pattern.

The DM   vector field $B_1$ of odd parity couples to the known fermion $f$
and its   KK parity-odd partner $F$ in the vertex $BfF$.
The Feynman amplitude for the process
$B(k,\mu) + f(p)  \to  B(\ell ,\nu) f(q) $ is
\be -C\bar u(q) \gamma^\nu
{\not p+\not k+ M_F \over
 (p +k)^2-M_F^2 }
\gamma^\mu u(p)
\ + \ \hbox{( crossing } \mu\leftrightarrow \nu \ ,
                        k  \leftrightarrow -\ell \ )
\ee
where $\mu$ and $\nu$ are the polarization indices.
The  heavy fermion $F$ of mass $M_F$ is exchanged
in the $s$ channel (the first term), and  in the $u$ channel
(the following parenthesis).
We take the heavy $M_F\gg p,q,\ell,k$ limit for which
\be (C/M_F^2)
\bar u(q)\left(
 \gamma^\nu(\not p+\not k)\gamma^\mu
+ \gamma^\mu(\not p-\not \ell)\gamma^\nu
+2M_F g^{\mu\nu} \right) u(p)   \ee
The $2M_F$ term flips chirality. If the KK mode interaction is chiral,
just like the SM, the term  $2M_F$ vanishes when chirality projections
are added in the amplitude.
We use the Chisholm identity,
\be
\gamma^\mu\gamma^\alpha\gamma^\nu
=g^{\mu\alpha}\gamma^\nu -g^{\mu\nu}\gamma^\alpha+g^{\alpha\nu}\gamma^\nu
+i\epsilon^{\mu\alpha\nu\beta}\gamma_\beta\gamma_5 \ ,\ee
and add the
$s$ and $u$ channels contributions,
\be  (C/M_F^2)
\bar u(q)\left(
(2p+k-\ell)^\nu \gamma^\mu + (2p+k-\ell)^\mu \gamma^\nu
-(2\not p +\not k-\not\ell) g^{\mu\nu}
+ i\epsilon^{\mu\alpha\nu\beta}(k+\ell)_\alpha \gamma_\beta\gamma_5 \right)
u(p)\ee
Then using the  4-momentum conservation, the Dirac equation of the massless
quark limit, and the physical polarization properties, we obtain
\be (C/M_F^2)
\bar u(q)\left(
(p+q)^\nu \gamma^\mu +(p+q)^\mu \gamma^\nu
+  i\epsilon^{\mu\alpha\nu\beta}(k+\ell)_\alpha \gamma_\beta\gamma_5
      \right)u(p) \ .\ee
The relevant operators in the chirality basis are
\be B_\mu B_\nu \
[{ \bar q }\
i  ( {\stackrel{\leftrightarrow}{\partial}}{\ }^\mu\gamma^\nu
 +  {\stackrel{\leftrightarrow}{\partial}}{\ }^\nu\gamma^\mu )
\hbox{$1\over2$}(1\pm\gamma_5) q]  \quad\hbox{ and }\quad
i\epsilon^{\mu\alpha\nu\beta}
(B_\nu i{\stackrel{\leftrightarrow}{\partial}}{\ }_\alpha  B_\mu )
\ { \bar q }\gamma_\beta (1\pm\gamma_5)    q \ . \ee
The SD interaction is given by
\be
{\cal O}_{\rm SD} =i\epsilon^{\mu\alpha\nu\beta}
(B_\nu i{\stackrel{\leftrightarrow}{\partial}}{\ }_\alpha  B_\mu )
\ { \bar q }\gamma_\beta\gamma_5    q  \ .\ee

\bibliographystyle{h-physrev}
\bibliography{dm-ident}                      


\end{document}